\documentstyle[psfig]{mn2e}

\newif\ifAMStwofonts

\def\simlt{\lower.5ex\hbox{$\; \buildrel < \over \sim \;$}}
\def\simgt{\lower.5ex\hbox{$\; \buildrel > \over \sim \;$}}

\def\mnras{MNRAS}
\def\apj{ApJ}
\def\apjl{ApJL}
\def\aap{AAp}

\title[The Evolution of Substructure II]
    {The Evolution of Substructure in Galaxy, Group \\ and Cluster Haloes II:
Global Properties}
\author[Taylor \& Babul]
{James E. Taylor$^{1}$\thanks{PPARC Fellow}\thanks{email: {\tt jet@astro.ox.ac.uk}} and Arif Babul$^{2}$ \\
$^{1}$Denys Wilkinson Building, 1 Keble Road, Oxford OX1 3RH, United Kingdom \\
$^{2}$Elliott Building, 3800 Finnerty Road, Victoria, BC, V8P 1A1, Canada \\}
\date{\today}
\pubyear{2004}

\begin{document}

\maketitle

\begin{abstract}
In a previous paper, we described a new method for including detailed
information about substructure in semi-analytic models of dark matter 
halo formation based on merger trees. In this paper, we present the 
basic predictions of our full model of halo formation. We first
describe the overall properties of substructure in galaxy, group or
cluster haloes at the present day. We then discuss the evolution of
substructure, and the effect of the mass accretion history of an 
individual halo on the mass function and orbital grouping of its subhalo
population. We show, in particular, that the shape of the subhalo
mass function is strongly correlated with the formation epoch of the
halo. In a third paper in this series we will compare the results of 
our semi-analytic method with the results of self-consistent numerical 
simulations of halo formation. 
\end{abstract}

\begin{keywords}
methods: numerical -- galaxies: clusters: general -- galaxies: formation -- galaxies: haloes -- dark matter.
\end{keywords}


\section{Introduction}\label{sec:1}

Following the analysis of the first-year data from the Wilkinson
Microwave Anisotropy Probe (WMAP), there is now very strong evidence
that the matter content of the Universe is in large part non-baryonic
(Spergel et al.\ 2003). Observations of large scale structure (e.g.\
Percival et al.\ 2001; Maller et al.\ 2003; Tegmark et al.\ 2004),
measurements of weak lensing (e.g.\ Hoekstra et al.\ 2002; Van
Waerbeke et al.\ 2002; Jarvis et al.\ 2003; Hamana et al.\ 2003;
Brown et al.\ 2003; Rhodes et al.\ 2004), and modelling of the
Lyman-$\alpha$ forest (e.g.\ Croft et al.\ 2002; Demia{\' n}ski \&
Doroshkevich 2003; Kim et al.\ 2004) indicate a power spectrum of
density fluctuations in this component continuing to subgalactic
scales, and consistent with `cold' dark matter (CDM) models. The
implications of the CDM power spectrum for structure formation are
well established. Dark matter haloes, the dense regions that surround
galaxies, groups and clusters, form from the bottom up, through the
merging of progressively larger structures. This process of
hierarchical merging has been studied extensively for two decades, and
the overall properties of galaxy or cluster haloes formed in this way
are now fairly well determined.

To learn more about dark matter, and to search for features in the
power spectrum that could reveal new phases in the evolution of the
very early universe, we must push the theory of structure formation
to smaller scales. Since the present-day evolution of structure is
already nonlinear on the scale of galaxy clusters, it is challenging
to make even the simplest predictions on subgalactic scales. The
usual approach to the problem is to combine the results of simulations
on different scales, using numerical simulations of large
(100\,Mpc$^{3}$ or more) volumes to determine the formation rates and
properties of the overall population of dark matter haloes at low
resolution (e.g.\ Kauffmann et al.\ 1999; Bullock et al.\ 2001;
Gottl{\" o}ber, Klypin, \& Kravtsov 2001; Yoshida, Sheth, \& Diaferio
2001; Evrard et al.\ 2002; Mathis et al.\ 2002;  Wambsganss, Bode \&
Ostriker 2003; Lin, Jing, \& Lin 2003; Gottl{\" o}ber et al.\ 2003;
Zhao et al.\ 2003a, 2003b; Hatton et al 2003; Reed et al.\ 2003;
Klypin et al.\ 2003; Percival et al.\ 2003; Tasitsiomi et al.\ 2004;
Yahagi, Nagashima, \& Yoshii 2004; Gao et al.\ 2004b), more detailed
re-simulations of objects selected from these volumes to determine the
structure of individual haloes for reasonably large samples (e.g.\
Klypin et al.\ 2001; Jing \& Suto 2002; Fukushige \& Makino 2001,
2003; Ascasibar et al.\ 2003; Hayashi et al.\ 2003; Fukushige, Kawai
\& Makino 2004; Hoeft, M{\" u}cket, \& Gottl{\" o}ber 2004; Navarro
et al.\ 2004; Tasitsiomi et al.\ 2004), and finally very
high-resolution simulations of individual haloes focusing on their
structure and substructure, either on cluster scales (e.g.\ Tormen 1997; 
Tormen, Diaferio \& Syer 1998; Ghigna et al.\ 1998;
Col{\'{\i}}n et al.\ 1999; Klypin 1999a; Okamoto \& Habe 1999;
Ghigna et al.\ 2000; Springel et al.\ 2001; Governato, Ghigna \&
Moore 2001; De Lucia et al.\ 2004; Gill, Knebe, \& Gibson 2004a;
Gill et al.\ 2004b; Gao et al.\ 2004a), on galaxy scales (e.g.\
Klypin et al.\ 1999b; Moore et al.\ 1999a, 1999b; Stoehr et al.\
2002; Power et al.\ 2003), or both (e.g.\ Diemand et al.\ 2004c; 
Gao et al.\ 2004b; Weller, Ostriker \& Bode 2004; Reed et al.\ 2004).

This approach has provided a detailed picture of the structure and
dynamics of dark-matter-dominated systems over a wide range of scales,
from tens of kiloparsecs to hundreds of Megaparsecs. There is a hard
limit to the dynamic range that can be achieved using a multi-scale
approach, however. Structure formation mixes information on many
different scales as haloes form. To model the formation of a halo
accurately, one needs to include the effects of very long-wavelength
fluctuations together with the smaller fluctuations that produce
substructure. Thus the minimum scale that can be included in any
self-consistent simulation of a present-day halo is constrained by 
the requirement that the total volume studied still be in the linear
regime (where the effect of larger fluctuations can be accounted for) 
at the present day, and by the finite numerical resolution
available computationally. For the highest-resolution simulations
possible currently, this leads to a minimum mass scale for resolved
substructure of around 10$^{-4}$--10$^{-5}$ of the mass of the main
halo considered. To study halo substructure below this mass limit
requires analytic or semi-analytic extensions to the numerical
results. It is precisely this sort of small-scale information,
however, that is required in many current applications including
galaxy dynamics, strong lensing, direct or indirect dark matter detection 
or tests of dark matter physics in general.

In earlier work (Taylor \& Babul 2001, TB01 hereafter), we developed a
model for dynamical evolution of satellites orbiting in the potential
of larger system. This model includes dynamical friction, tidal mass
loss and tidal disruption as its main components. It calculates
satellite evolution over a many short timesteps, rather like a
restricted $N$-body simulation, but uses only global properties of
the satellite to determine its evolution, thus reducing the
computational expense considerably. More recently (Taylor \& Babul
2004a, paper I hereafter), we applied this model of satellite
evolution to the merging subcomponents involved in the hierarchical
formation of galaxy, group or cluster haloes. As one of the main
results of paper I, we established a non-parametric way to include
realistic amounts of substructure in semi-analytic merger trees
without increasing the computational cost substantially. This
`pruning' method, together with established merger tree methods and
the analytic model of satellite dynamics from TB01, constitute a new
semi-analytic model for the formation of individual dark matter haloes.

In this paper, we discuss the basic predictions of this model. In section 
\ref{sec:2}, we first review the model, which was described fully in paper I. 
In section \ref{sec:3}, we then discuss the distribution functions for the  
mass, peak  circular velocity, and spatial location of subhaloes that it 
predicts. In section \ref{sec:4}, we show that these results are relatively 
insensitive to specific model parameters, uncertainties or assumptions. 
The model predicts strong correlations among the different properties of
subhaloes, and between the age of subhaloes and their other properties, 
which are discussed in section \ref{sec:5}. In section \ref{sec:6}, we show 
how the systematic dependence of subhalo properties on age may allow the use  
of substructure as an indicator of the dynamical history of an individual  
system. Finally in section \ref{sec:7}, we consider kinematic grouping of 
substructure due to the merger sequence, and the effects of encounters  
between subhaloes. We discuss our results in section \ref{sec:8}.

In a subsequent paper (Taylor \& Babul 2004b, paper III hereafter) we
will compare the predictions of this semi-analytic model with the
results of self-consistent numerical simulations of halo
formation. This comparison is  particularly interesting, since the
only free parameters in the semi-analytic  model were fixed in paper
I, either by matching restricted simulations of  individual subhaloes
(to fix the parameters of the dynamical model), or by  assuming
self-similarity in the merging process (to fix the one parameter in
the pruning method). Thus we have no remaining parametric freedom when
comparing our results to self-consistent simulations, making the
comparison a meaningful one.

Finally, we note that as in paper I, here and in paper III we will 
generally consider results for the former `standard' CDM (SCDM) cosmology 
with $h = 0.5$ and $\sigma_8 = 0.7$, because the simulations we compare 
to directly in paper III assumed this cosmology. As illustrated in section 
\ref{subsec:4.2}, the properties of substructure depend only 
weakly on cosmology, and thus we expect our main results to be equally
valid for $\Lambda$CDM or similar cosmologies. We will consider the 
cosmological dependence of our results in more detail in future work.

\section{Review of the Semi-analytic Model}\label{sec:2}

In paper I, we described a full semi-analytic model for studying the
formation of dark matter haloes and the evolution of their
substructure. In this section we will review briefly the main
features of this model. The semi-analytic model consists of several
components: a method for generating merger trees, an algorithm for
`pruning' these trees, to determine how many distinct satellites merge
into the main system within the tree, and an analytic model to
describe the subsequent evolution of these satellites.

The merger tree method is an implementation of the algorithm proposed
by Somerville and Kolatt (1999). It decomposes a present-day halo into
its progenitors at earlier times, using small timesteps in order to
preserve Press-Schechter (Press \& Schechter 1974) statistics. The
merger trees considered in this paper generally have a total mass of
$1.6\times10^{12}M_{\odot}$ at $z = 0$ and a mass resolution of
$5\times10^{7}M_{\odot}$, chosen to match the simulations described 
in paper III; below this resolution limit they become
increasingly incomplete. The merger tree is traced back until all
branches drop below the mass resolution limit or a redshift of 30 is
reached, and typically contains tens of thousands of branchings.

Within each tree, we can define a main `trunk' by tracing the merger
history of the final system back from the present day, choosing the
most massive progenitor every time the tree branches, and following
this object back in redshift. We will also refer to this as the `main
system' or the `main halo' in the tree. We will describe branches off
the main trunk as `first-order mergers', branches off these branches
as `second-order mergers', and so on. To avoid the computational cost
of following the evolution of each sub-branch in the tree, we consider
in detail only those systems that merge with the main trunk of the
tree. To treat higher-order branches, we use a simplified description
of satellite dynamics to `prune' the merger tree, as described in
paper I. We assume that the dynamical age of a satellite, or more
specifically the number of orbits it has spent in a system, determines
whether it is loosely bound and should be treated as a distinct object
when its parent merges with a larger halo, or whether it is tightly
bound and should be considered part of its parent. By self-similarity,
we established in paper I that the average time delay separating the
two cases should be $n_{\rm o} \simeq$ 2.0--2.2 orbital periods,
depending on the disruption criterion assumed. Thus if higher-order
substructure has spent more than 2 orbits in its parent halo when that
parent merges with a lower-order branch, then we treat it as part of
its parent. Otherwise, we treat it as a distinct merger with the
lower-order system.

As a satellite evolves in a parent's halo, it also loses mass. From
self-similarity, we determined in paper I that satellites that had
spent $n_{\rm o}$ orbital periods or less in their parent system
retained an average fraction $f_{\rm st} = 0.73$ of their original
mass. Thus if a satellite of mass $M_{\rm s}$ has spent less than
$n_{\rm o}$ orbits in its parent's halo, it is passed on to the next
lower-order system with a mass $0.73\,M_{\rm s}$, while the remaining
mass is added to its original parent. In the full merger tree,
satellites of initial order $n$ will thus be reduced to 
$(f_{\rm st})^{(n - 1)}$ of their original mass when they finally merge 
with the main trunk. (Here and in what follows, the `original mass'
$M_{\rm s,0}$ is the mass a subhalo has before it merges with a larger
system for the first time, while the `infall mass' $M_{\rm s,i}$ is
the mass a subhalo has when it merges with the main trunk.)

Using this `pruning' method, we determine how much substructure
percolates down to the main trunk of the merger tree. After being
pruned in this way, a typical merger tree contains roughly 4000
branches, each recording the merger of a single satellite with the
main system. If these satellites were part of the same branch before
pruning, we consider them to be part of the same kinematic group, and
give them correlated initial orbits, as described in paper
I. Otherwise they are distributed on randomly oriented orbits starting
at the virial radius, with a realistic distribution of angular
momenta. The only other property we need to specify for the subhaloes
is their initial density profile, which is assumed to be a Moore
profile, $\rho (r) = \rho_0 r^{1.5}/(r_{\rm s}^{1.5} + r^{1.5})$. 
Although the most recent high-resolution simulations (Power
et al.\ 2003; Fukushige, Kawai \& Makino 2003; Navarro et al.\ 2004;
Tasitsiomi et al.\ 2004; Diemand et al.\ 2004b) suggest that typical
haloes have a central density profile with a slope closer to the value
of $-1$ suggested by Navarro, Frenk and White (1996, 1997, NFW
hereafter), the Moore fit gives a better representation of the excess
central mass, relative to the NFW fit, seen in the simulated haloes.

The subhalo profiles can also be reparameterised in terms of their
original virial radius $r_{\rm vir}$\footnote{We will use the
standard definition of $r_{\rm vir}$ derived from the spherical
collapse model, as the radius within which an isolated halo has a
mean density $\rho_{\rm vir}$ that exceeds the critical density
$\rho_{\rm c}$ by a factor $\Delta_{\rm c}$ which depends on
cosmology.} and their concentration $c \equiv r_{\rm vir}/r_{\rm s}$.
Since the virial radius of a halo depends only on its total mass and
on the background cosmology, isolated spherical haloes of a given
mass at a given redshift are completely specified by their
concentration. We determine concentrations by applying the relations
of Eke, Navarro and Steinmetz (2001, ENS01 hereafter) to each system
at the time it {\it first appears} in the merger tree, that is the
last time it is an independent object, as opposed to substructure in
a larger system. (We divide the ENS01 concentration by a factor
$r_{\rm s,N}/r_{\rm s,M} = 1.73$, to account for the relative
difference in scale radius between the Moore and NFW profiles, as
explained in paper I.) If the subhalo is stripped before falling into
the main system, its density profile is modified according to the
formula proposed in Hayashi et al.\ (2003, H03 hereafter; equation 8),
specifically:
\begin{equation}
\rho (r) = {{f_{\rm c}}\over{1 + ({r}/{r_{\rm c}})^3}}\,\rho_0(r)\, ,
\end{equation} 
where $\rho_0$ is the original density profile, $f_{\rm c}$ describes
the reduction in central density as the system is stripped, and
$r_{\rm c}$ indicates the radius beyond which the system is almost
completely stripped by tides. Formulae relating $f_{\rm c}$, $r_{\rm c}$ 
and the declining peak circular velocity to the net amount of
mass loss are given in H03 (see their Fig.\ 12).

The subsequent evolution of the individual subhaloes is determined
using the analytic model of TB01 to account for dynamical friction,
mass loss, tidal heating and disruption. The evolution of the density
profile as mass is lost is described in H03. There is some
uncertainty as to when satellites will be completely disrupted by
repeated mass loss. As discussed in paper I, we consider satellites
disrupted if they are stripped down to a critical fraction of their
mass, $f_{\rm dis}$ corresponding to the mass within either 0.5 or
0.1 times the binding radius defined by H03. (For the satellites
considered here this is equivalent to 2--4 percent or 0.2--0.4
percent of the original mass, respectively.) In what follows, we will
refer to these two cases as models `A' and `B'. Satellites are also
considered disrupted if they enter into the central $0.01 r_{\rm vir,m}$ 
of the main halo, since we expect mass loss to be rapid in
this region, but we cannot follow satellite evolution accurately this
far into the potential without using much shorter timesteps than are
required for the rest of the tree.

The basic properties of the main system, specifically its mass and
virial radius at a given redshift, are also determined directly from
the merger tree. We assume unless specified otherwise that the main
system has a Moore density profile, and a concentration or scale
radius given by the relations in ENS01. Our fiducial system, a
$1.6\times10^{12}M_{\odot}$ halo at $z = 0$ in a SCDM cosmology, has a
concentration $c_{\rm M} = 10.3$, a scale radius $r_{\rm s,M} =
30.5$\,kpc, a virial radius $r_{\rm vir,m} = 314.1$\,kpc, and a
virial velocity (or circular velocity at the virial radius) 
$v_{\rm vir} = 148$\,km\,s$^{-1}$. We note that this concentration
is typical for a galaxy of this mass (ENS01); galaxy clusters would
be about half as concentrated, so one must keep this difference in mind
when comparing our results with simulations of more massive systems. 
On the other hand, real galaxy haloes have large concentrations of
baryonic material at their centres, and through adiabatic contraction 
they may have become more concentrated than the systems considered here; 
this possible difference should be kept in mind when comparing with
observations. 

In all, the dynamical model has two main free parameters -- the Coulomb
logarithm $\Lambda_{\rm s}$ which modulates dynamical friction, and
the heating coefficient $\epsilon_{\rm h}$ which modulates mass
loss. (A third parameter discussed in TB01, the disk logarithm
$\Lambda_{\rm d}$, is not used here since we are considering
evolution in a single-component potential). The precise disruption
criterion (say the fraction of the binding radius used to define
$f_{\rm dis}$), the form chosen for the density profile of the
satellites and the profile of the main system, and various other model
choices will also affect some of our results, though not very
strongly. We will discuss the model-dependence of our results in
section \ref{sec:4}. We first present results for the default
parameter values discussed in paper I, specifically 
$\Lambda_{\rm s} = 2.4$ (where the magnitude of dynamical friction 
scales as $\Lambda (M) = \Lambda_{\rm s} + \ln(M_{\rm h}/140\,M_{\rm s})$ 
if $m < M/140$, and $\Lambda (M) = \Lambda_{\rm s}$ for $m \ge M/140$), 
and $\epsilon_{\rm h} = 3.0$. The disruption criterion assumes either
$f_{\rm dis} = 0.5$ (model A) or $f_{\rm dis} = 0.1$ (model B). Given
these parameter choices, the pruning parameters are fixed iteratively
as discussed in paper I. For model A $n_{\rm o} = 2.0$, and for model
B $n_{\rm o} = 2.2$, while $f_{\rm st} = 0.73$ in both cases.

\section{Basic Predictions for Halo Substructure at $z = 0$}\label{sec:3}

\subsection{Cumulative distributions and higher moments}\label{subsec:3.1}

The first high-resolution simulations of galaxy haloes (Klypin et al.\
1999b; Moore et al.\ 1999a) demonstrated that halo substructure is
very close to self-similar across a wide range of halo mass and for
different CDM cosmologies. In particular, the cumulative number of
subhaloes with internal circular velocities equal to a given fraction
$f$ of the velocity of the main halo, as a function of $f$, is
roughly universal. A number of other simulations (e.g.\ Okamoto \&
Habe 1999; Springel et al.\ 2001; Governato et al.\ 2001; Stoehr et
al.\ 2002; Desai et al.\ 2004; Reed et al.\ 2004; De Lucia et al.\
2004; Gao et al.\ 2004b; Weller et al.\ 2004) have confirmed the
universal form for the cumulative distribution of relative circular
velocity (or `cumulative relative velocity function'), and have shown
that the cumulative relative mass function is also roughly
universal. Thus we will quantify the predictions of the semi-analytic
model in terms of these cumulative functions $N(>v_{\rm p})$ and
$N(>M)$, as well as the higher-order dependence of these functions on
position within the halo. We re-examine the universality of the
cumulative functions derived from numerical simulations in paper III.

\subsubsection{$N(> M)$}\label{subsubsec:3.1.1}

Although there are different conventions for determining the total
mass of a halo or subhalo, they yield similar values, and therefore
the normalisation of the cumulative mass function is
straightforward. We choose the convention of normalising subhalo
masses (denoted $M_{\rm s}$) to the virial mass of the main system
predicted by the merger tree model at the redshift considered
(normally $z = 0$), $M_{\rm vir,m}$. As explained in paper I, this is
the mass within a region of mean density $\Delta_{\rm c}\rho_{\rm c}$
at that redshift. Since our fiducial mass $1.6\times10^{12}M_{\odot}$
is a reasonable estimate for the total virialized mass surrounding a
system like the Milky Way at $z = 0$, we will also indicate the
absolute values of subhalo masses, to facilitate comparison with the
satellites of the Milky Way. We will discuss the normalisation of the
numerical results in section paper III.

\begin{figure}
 \centerline{\psfig{figure=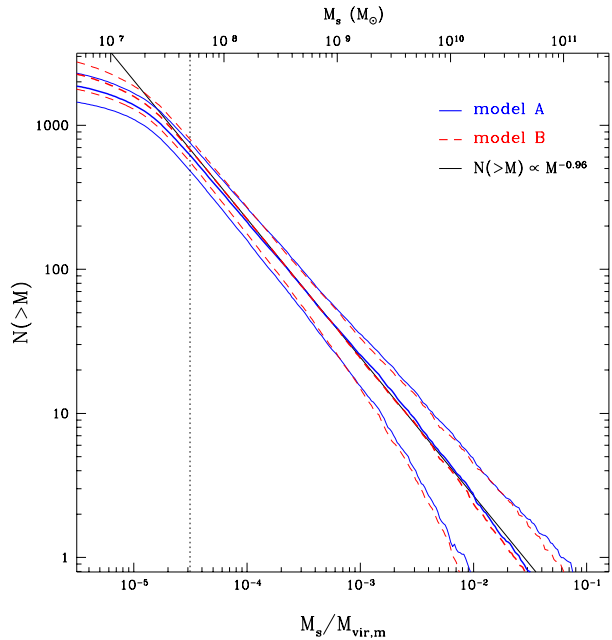,width=1.0\linewidth,clip=,angle=0}}
 \caption[]{The cumulative mass functions predicted by the
semi-analytic model. The thick lines show the average result for a
104 SCDM merger trees at $z = 0$, for model A (solid lines) and model
B (dashed lines). The thin lines show the 1-$\sigma$ halo-to-halo
scatter. The thin solid line shows a power-law of slope -0.96, while
the vertical line indicates the mass resolution limit of the original
merger trees. }\label{fig:1}
\end{figure}

Fig.\ \ref{fig:1} shows the cumulative relative mass functions
predicted by the semi-analytic model at $z = 0$. The thick lines are
the mean for a set of 104 trees, and the thin lines indicate the
1-$\sigma$ halo-to-halo scatter. (Solid lines are for model A and
dashed lines are for model B.) The top axis indicates the
corresponding subhalo mass in a system with a virial mass of
$1.6\times10^{12}M_{\odot}$. Over the range resolved by the merger
trees ($M_{\rm s} > 5\times10^{7}M_{\odot}$), the cumulative mass
function is very close to a power law: 
$N(>M) = N_0 (M_{\rm s}/M_{\rm vir,m})^{\alpha}$, with $N_0 = 0.038$ 
and $\alpha = -0.93$ for model
A and $N_0 = 0.032$ and $\alpha = -0.96$ for model B (this fit is
included as a thin solid line on the plot). The 1-$\sigma$
halo-to-halo scatter is approximately 20--30 percent at the low mass
end, but more than a factor of two at the high mass end.

\subsubsection{$N(>v_{\rm p})$}\label{subsubsec:3.1.2}

\begin{figure}
 \centerline{\psfig{figure=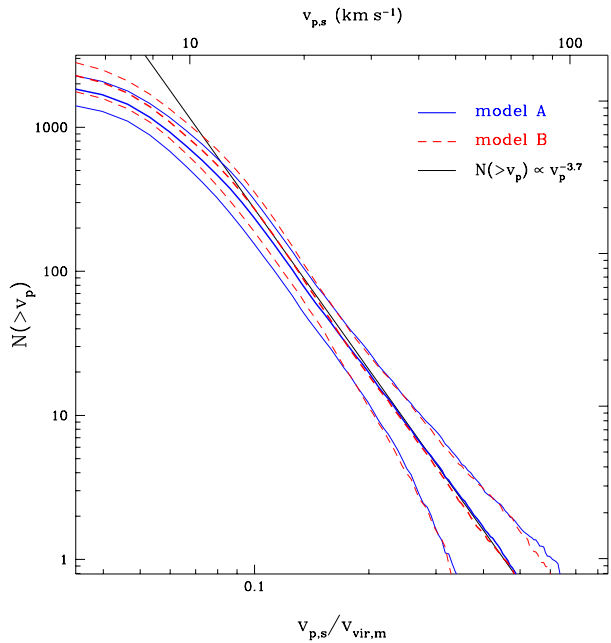,width=1.0\linewidth,clip=,angle=0}}
 \caption[]{The cumulative peak circular velocity functions predicted
by the semi-analytic model. The thick lines show the average result
for a 104 SCDM merger trees at $z = 0$, for model A (solid lines) and
model B (dashed lines). The thin lines show the 1-$\sigma$
halo-to-halo scatter. The thin solid line shows a power-law of slope
-3.7. }
\label{fig:2}
\end{figure}

The normalisation of velocities is more problematic. Two different
velocities, the virial velocity (or circular velocity at the virial
radius) $v_{\rm vir} = \sqrt{G\,M_{\rm vir}/r_{\rm vir}}$, and the
velocity at the peak of the rotation curve $v_{\rm p}$, can be used
to characterise an isolated halo\footnote{For an NFW profile of
concentration $c$, the ratio 
$v_{\rm p}/v_{\rm vir} = 0.465[\ln(1+c)/c - 1/(1+c)]^{-0.5}$. 
The approximation $v_{\rm p}/v_{\rm vir} = 1.0 + 0.025(c - 2.163)$ 
holds to better than 2 percent for $c =$ 2--20, the
typical range for galaxy and cluster haloes.}. Subhaloes within a
larger system will be tidally truncated, so the equivalent of the
virial velocity in these systems will be the circular velocity at some
outer radius that marks their tidal limit. In order to compare our
the properties of subhaloes to simulations, we will consider the peak
velocity $v_{\rm p}$, as it is easier to measure reliably in
simulations, whereas the outer radius of a system may be poorly
determined. On the other hand, the peak velocity has an additional
and quite large dependence on the concentration of the system. Thus,
we will normalise subhalo velocities to the virial velocity of the
main system (denoted $v_{\rm vir,m}$) rather than its peak velocity
when comparing results for different haloes, and plot distributions
in terms of the velocity ratio $v_{\rm p,s}/v_{\rm vir,m}$.

Fig.\ \ref{fig:2} shows the cumulative relative velocity function
predicted by our model at $z = 0$, that is the cumulative distribution
of peak circular velocity $v_{\rm p,s}$, relative to the virial
velocity $v_{\rm vir,m}$ of the main system. Line styles are as in
Fig.\ \ref{fig:1}, and as before the top axis indicates the
corresponding peak velocities of subhaloes in a system with a virial
velocity of 148\,km\,s$^{-1}$. In the range resolved by the merger
trees ($v_{\rm p,s} \simgt 10$\,km\,s$^{-1}$), the cumulative
velocity function is close to a power law: 
$N(>v_{\rm p,s}) = N_0(v_{\rm p,s}/v_{\rm vir,m})^{\beta}$, 
with $N_0 = 0.07$ and $\beta = -3.5$ 
for model A and $N_0 = 0.054$ and $\beta = -3.7$ for model B
(this fit is included as a thin solid line on the plot). We note that
if all subhaloes had peak velocities proportional to their original
virial velocities, we would expect a slope $\beta = 3\,\alpha$. the
actual slope is steeper due to the concentration-mass-redshift
relation, which implies that low-velocity systems are more
concentrated on average, and thus that have higher values of 
$v_{\rm p}/v_{\rm vir}$ initially. Finally, we note that as with the 
mass function, the 1-$\sigma$ halo-to-halo scatter in the velocity
function is 20--30 percent at the low velocity end, but more than a
factor of two at the high-velocity end.

\subsubsection{$N(M \| R)$}\label{subsubsec:3.1.3}

\begin{figure}
 \centerline{\psfig{figure=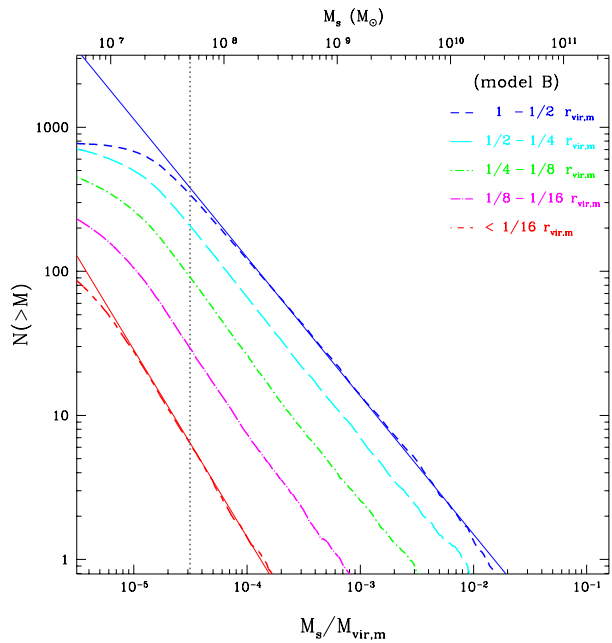,width=1.0\linewidth,clip=,angle=0}}
 \caption[]{The cumulative mass function of subhaloes in different
radial bins, for haloes generated using model B. The slope of the mass
function varies from $\alpha = -0.96$ in the outermost radial bin to
$\alpha = -1.3$ in the innermost bin, as indicated by the thin lines.
}
\label{fig:3}
\end{figure}

We expect the mass and circular velocity of a subhalo to be correlated
with its position within the main halo. This is both because subhaloes
close to the centre of the main system will have merged with the main
system at a higher redshifts on average, and also because they will
have lost more mass through tidal stripping. Fig.\ \ref{fig:3} shows
how the cumulative mass function varies, when binned in radius. In
the outer parts of the halo, it has essentially the same slope as the
mass function for the whole halo, $\alpha = -0.96$. In the innermost
regions, the slope steepens, so that within $r < r_{\rm vir,m}/16$ 
($\simeq $ 20\,kpc in our fiducial system), $\alpha = -1.3$. Since the
differential mass function goes as $dN/dM \propto M^{\alpha -1}$, and
the contribution to the mass from substructure of mass $M$ or greater
goes as $f(>M) \propto M^{(\alpha -1) + 1} = M^{\alpha}$, a slope of
$\alpha < -1$ implies that this quantity will diverge as we integrate
down to smaller and smaller masses. Thus, the question of how much of
the dark matter at the centres of haloes is in bound substructure
remains highly uncertain. This uncertainty has important implications
for attempts to detect dark matter observationally or experimentally,
as discussed in paper III.

\subsubsection{$N(R \| M)$, $N(R \| v_{\rm p})$}\label{subsubsec:3.1.4}

\begin{figure*}
 \centerline{\psfig{figure=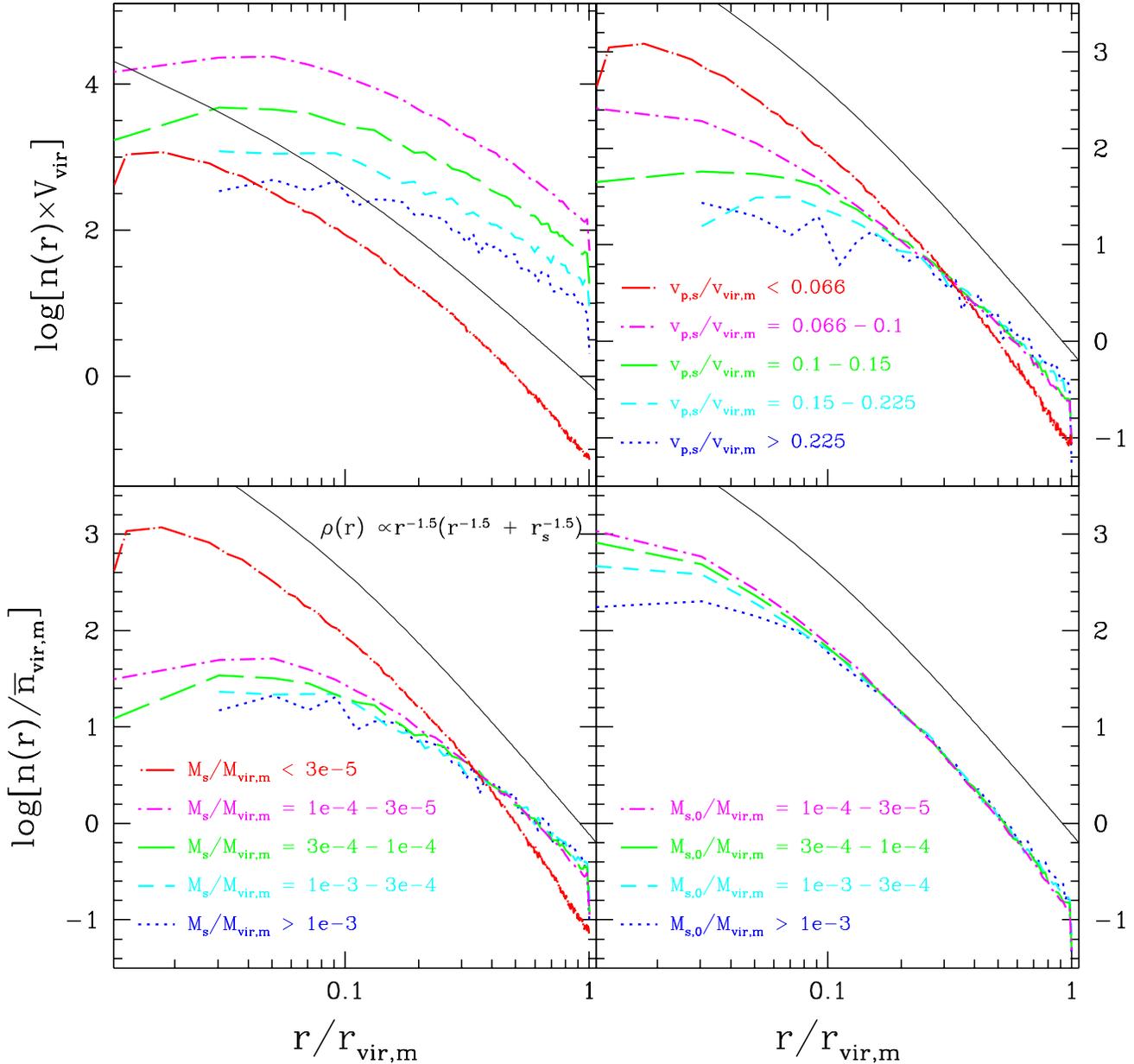,width=1.0\linewidth,clip=,angle=0}}
 \caption[]{The number density of subhaloes as a function of their
position in the main system, binned and normalised in various
ways. The top left-hand panel shows the number density multiplied by
the virial volume, binned by their mass at $z = 0$. The other panels
show the number density normalised to the mean within the virial
radius for objects in that bin. Systems are binned by their mass at
$z = 0$ (bottom left), their original mass (as defined in the text --
bottom right), and by their peak circular velocity at $z = 0$ (top
right). In each case, only systems that survive at $z = 0$ are
included. The thin lines show a Moore density profile of concentration
$c_{\rm M} = 10$, with arbitrary normalisation. }
\label{fig:4}
\end{figure*}

We can also take other moments of the substructure distribution.
Fig.\ \ref{fig:4} shows the radial number density profile of the
subhalo distribution, for subhaloes binned in several different ways.
In the top left panel, the different curves all have the same
normalisation (they are all multiplied by the volume within the
virial radius, ${\rm V}_{\rm vir} = 4\pi\,r_{\rm vir,m}^{3}/3$). In
the other panels, the number density in each bin has been normalised
to the mean number density within the virial radius for objects in
that mass range. Systems are binned by their mass at $z = 0$ (two
left-hand panels), their `original' mass $m_{s,0}$ defined previously
(bottom right panel), and by their peak circular velocity at $z = 0$
(top right panel). In each case, only systems that survive at $z = 0$
are included. The thin lines show a Moore density profile of
concentration $c_{\rm M} = 10$, with arbitrary normalisation.

In the left-hand panels, in the four mass bins above our resolution
limit ($5\times 10^{7}M_{\odot}$, or $3\times 10^{-5}\,M_{\rm vir,m}$), 
the number density profiles are appreciably shallower than
the background density profile. They appear to have central cores of
roughly constant number density, ranging in size from 5 to 15 percent
of the virial radius in size (15--45\,kpc in our fiducial system),
depending on the mass bin. From the absolute scaling in the top
left-hand panel, it is clear that the total distribution summing over
the four bins above the resolution limit will be dominated by the
subhaloes in the lowest mass bin, and will therefore have a profile
similar to the lower dot-dashed line. Given the steepness of the
subhalo mass function, this will always be the case for number
weighted-distributions, and thus the radial distributions determined
in simulations will depend on the limiting mass resolution and the 
completeness near that limit. This may be one reason why different
methods for identifying substructure can determine quite different
radial distributions for the same simulation (Gill et al.\ 2004a, Fig.\ 9).

Objects in the mass bin below the resolution limit have a radial
distribution that is even {\it more} centrally concentrated than the
background density profile. This result is mainly due to the
incompleteness of the merger tree at these masses, however. Given our
initial mass resolution, the systems in this bin must have lost some
mass, and most will have lost a large fraction of their mass. This
explains why they are preferentially found in the central part of the
halo where mass loss rates are higher. Extending the merger tree
completely down to this mass range would reveal more low-mass systems
in the outer part of the halo, producing a shallower
profile. Overall, the trend in the radial distributions above the
mass resolution limit suggest that the subhalo distribution will
approach the background mass distribution as the resolution limit
decreases. The convergence to this final profile may be very slow,
however, as discussed in paper III.

The two right-hand panels show the normalised distributions of
subhaloes binned in various other ways. The top right-hand panel shows
the normalised distribution of systems binned by present-day peak
circular velocity, since this may be more relevant in comparisons
with observational data (Nagai \& Kravtsov 2004). Unfortunately the
resolution limit of the tree is not defined precisely in this
variable, so all the bins below $v_{\rm p,s}/v_{\rm vir,m} \sim 0.1$
are affected to some degree by the incompleteness of the merger tree.
Otherwise, the general trends are as in the bottom left-hand panel.

The bottom right-hand panel shows all surviving systems binned in
terms of their original mass. Down to 0.1 $r_{\rm vir,m}$, there is
very little difference in the results as a function of mass. This
agrees with the numerical results of Nagai \& Kravtsov (2004), who
find that there is little variation in the radial distribution of
subhaloes as a function of their mass at the time of accretion onto
the main halo. The overall distribution matches the mass distribution
of the main system very closely, as found by Nagai \& Kravtsov (2004)
and Gao et al.\ (2004), who used semi-analytic galaxies to trace the
distribution of stripped haloes in their cluster simulations. In the
innermost regions of the halo, we do find fewer surviving remnants of
massive subhaloes, relative to less massive subhaloes. This is partly
because the subhaloes in these regions are older, and the average mass
scale of substructure was smaller at earlier times, but it is also
because dynamical friction has caused many of the more massive central
systems to fall in to the centre of the potential and be disrupted.

\section{Model Dependence}\label{sec:4}

\subsection{Dependence on model parameters}\label{subsec:4.1}

Since the relative efficiency of various physical processes is
parameterised explicitly in the semi-analytic model, we can test the
sensitivity of our results to the input physics by running the model
with different values for these parameters. In particular, we can
test how the amplitude of the mass function depends on 
$\epsilon_{\rm h}$, $\ln\Lambda_{\rm h}$ and other parameters in the 
model. The
variation in the results is generally fairly small, so to highlight
the differences, we take parameter values at the extremes of, or even
beyond, their probable range, and normalise the results for these
variant models to the fiducial results from model B.

\begin{figure}
 \centerline{\psfig{figure=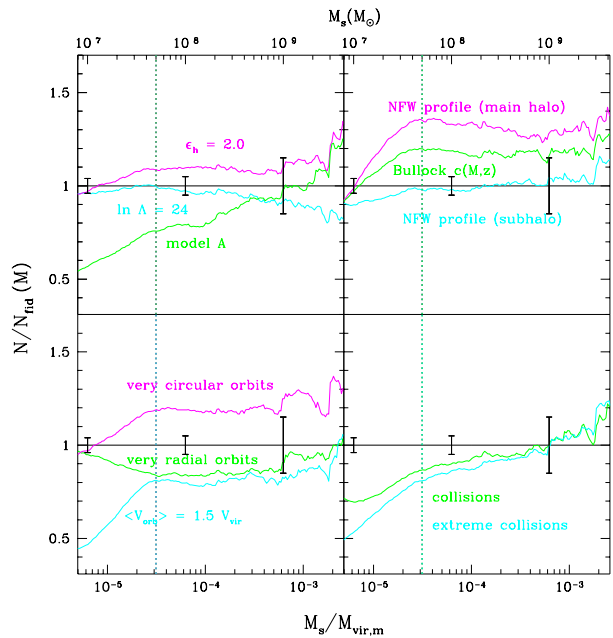,width=1.0\linewidth,clip=,angle=0}}
 \caption[]{The relative change in the (differential) mass function
for trees evolved in different variants of the basic model, including
(clockwise from upper left): a) varying the parameters describing
subhalo dynamics (model A is also shown for comparison), b) varying
the density profiles of the subhaloes or the main halo c) including
additional mass loss and disruption due to collisions, or d) varying
the distribution of subhalo orbits. In each case, we divide the
variant mass function by the fiducial mass function. The error bars
show the approximate level of 1-$\sigma$ fluctuations in the fiducial
mass function. The vertical dotted line indicates the resolution
limit of the merger trees; note that many of the mass functions change
systematically below this limit due to incompleteness. }\label{fig:5}
\end{figure}

Fig.\ \ref{fig:5} shows the amplitude of the (differential) subhalo
mass function for different variant models, relative to the mass
function in the fiducial model. These are (clockwise from the upper
left panel) a) different values for the parameters describing subhalo
dynamics, specifically $\epsilon_{\rm h} = 2.0$ rather than 3.0 or
$\ln\Lambda_{\rm h} = 24$ rather than 2.4 (model A is also shown for
comparison);  b) using an NFW density profile for the main halo,
using NFW profiles for the subhaloes, or using the concentration
relation from Bullock et al.\ (2001); c) including additional mass
loss and disruption due to collisions (see section \ref{subsec:7.2});
and d) using very circular ($\overline{\epsilon} = 0.6$ rather than
0.4) or very radial distributions ($\overline{\epsilon} = 0.25$ and
$\sigma_{\epsilon} = 0.17$) for subhalo orbits, or increasing the
initial orbital velocity to $1.5 v_{\rm vir,m}$. In each case, we
divide the variant mass function by the fiducial mass function. The
error bars show show the approximate level of 1-$\sigma$ fluctuations
in the fiducial mass function. Note that we have not rerun our full
model to determine a consistent pruning condition for each of these
variant runs; thus all we show here is how strongly the model
parameters affect a given set of subhaloes.

Overall, the mass function appears to be remarkably robust to changes
in the model, particularly since we have chosen parameter values that
are actually outside the expected range of uncertainty in many cases,
in order to produce a clear effect on the mass function. Increasing
the strength of dynamical friction reduces the number of massive
satellites that survive in the system, but we have had to increase
the Coulomb logarithm by a factor of ten to see a clear difference in
the mass function; for realistic values of $\ln \,\Lambda$ the effect
of any uncertainty in the exact strength of dynamical friction is
negligible. This is probably because dynamical friction is relatively
ineffective much below $M_{\rm s}/M_{\rm vir,m} = 10^{-2}$ (cf.\
paper I), and the average system has few satellites in this range.
Heating has a slightly larger effect; reducing the heating coefficient
by 30 percent, a reasonable change given the uncertainty in its
value, increases the normalisation of the mass function by roughly 10
percent. The difference between models A and B is also 10--20 percent
in amplitude; changing the pruning parameters $n_{\rm o}$ or 
$\Delta M_{\rm st}$ by 10 percent produces a change of similar magnitude,
steepening the slope of the mass function as $n_{\rm o}$ increases or
$\Delta M_{\rm st}$ decreases.

Changing the subhalo density profile to an NFW profile with a
shallower central cusp has almost no effect on the mass function (top
right-hand panel). If we use the concentration relations of Bullock
et al.\ (2001), small haloes are more concentrated and less sensitive
to tidal stripping, increasing the amplitude of the mass function by
20 percent. Changing the profile of the main system has a larger
effect however, increasing the amplitude of the mass function by 30
percent. Increasing the concentration of the main halo sufficiently
also produces similar results. This demonstrates the importance of
the central cusp in stripping and disrupting subhaloes as they pass
through the pericentres of their orbits. It also implies that adding
a central galactic component to the halo may reduce considerably the
number of subhaloes of a given mass that survive in the central part
of a halo. (We note that the numerical simulations analysed in
paper III are dark-matter only, so their lack of central
substructure is not be due to the disruptive effects of a central
galaxy.)

One process that is not included in our basic model of subhalo
dynamics is the effect of collisions and encounters between
subhaloes, or `harassment'. In section \ref{subsec:7.2} we will
describe an approximate way of recording collisions and modelling
their contribution to mass loss and disruption. The bottom right-hand
panel shows that the overall effects predicted by this model are
fairly minor, however; the amplitude of the mass function is reduced
at intermediate and small masses, but only by 10--15 percent (the
overall slope of the mass function also changes systematically,
decreasing by about 10 percent). Thus collisions do not seem to offer
a plausible way of getting rid of substructure.

Finally, the lower left-hand panel shows the effects of varying
subhalo orbits. Increasing the circularity of subhalo orbits (from
$\overline{\epsilon} = 0.4$ to $\overline{\epsilon} = 0.6$) produces
has a reasonably large effect, presumably because it moves their
pericentres out of the centre of the system. Khochfar and Burkert
recently reported an initial distribution of subhalo orbits from
large-scale simulations that was more {\it radial} than assumed
here. Although it has since become apparent that the radial bias was
due to an unsubtracted contribution from the Hubble flow (Khochfar
and Burkert 2003; Benson 2004), we can nonetheless test what effect
this sort of distribution would have on our results. To match the
distribution of circularities they first reported, we take a Gaussian
distribution with $\overline{\epsilon} = 0.25$ and 
$\sigma_{\epsilon} = 0.17$. This results in a 15 percent reduction 
in the amplitude of
the mass function. Finally, the third curve in this panel shows the
effect of varying initial subhalo velocities. The theoretical
uncertainty in this quantity turns out to be less important, as even
an unrealistically large change (increasing the mean velocity by 50
percent, or more than doubling the mean kinetic energy) produces only
a 20 percent reduction in the amplitude of the mass function.

In summary, testing our model for sensitivity to uncertainties in its
different components, we find that variations in the model parameters
over a plausible range have only a minor effect on the subhalo mass
function. In particular, given reasonable variations in the
parameters, the mass function is very insensitive to the exact
magnitude of dynamical friction, the central slope of the subhalo
density profile, or to minor collisions. It is mildly sensitive
(varying by 10--15 percent for reasonable variations in the
parameters) to subhalo concentrations, more radial subhalo orbits,
larger subhalo velocities or strong collisions. Only changing the
central density profile of the main halo or placing subhaloes on very
circular orbits has a large (20--30 percent) effect on the mass
function. This is consistent with the general conclusion from paper I
that a subhalo's properties are strongly affected by the density of
the main system at the pericentre of its orbit, through friction,
stripping, and disruption. Other subhalo properties, such as their
spatial distribution, may also vary systematically with the input
parameters. We discuss this variation in section paper III.

\subsection{Dependence on cosmology}\label{subsec:4.2}

\begin{figure}
 \centerline{\psfig{figure=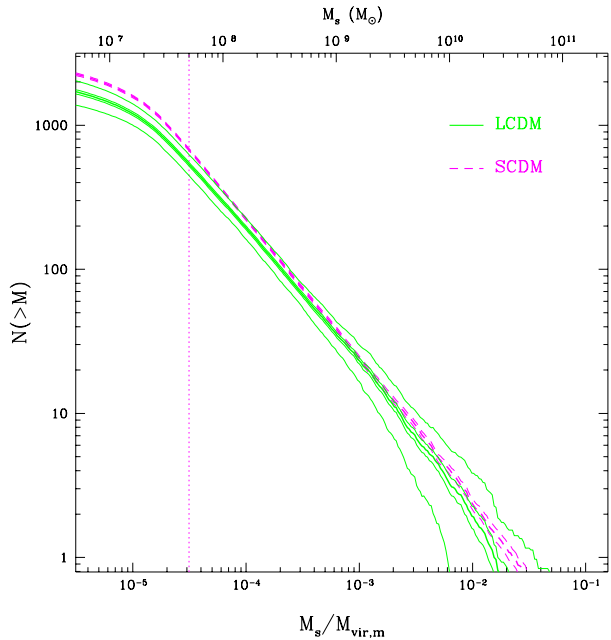,width=1.0\linewidth,clip=,angle=0}}
 \caption[]{The average cumulative mass function for LCDM subhaloes
(thick solid line, with thin lines indicating the 1-$\sigma$
uncertainty in the mean and the 1-$\sigma$ halo-to-halo scatter),
compared with the SCDM mass function (dotted lines, with upper and
lower lines showing the 1-$\sigma$ uncertainty in the mean).}
\label{fig:6}
\end{figure}

In most of this paper, we will present results generated in an SCDM
cosmology, in order to compare them with earlier numerical work which
assumed this cosmology. There are now much stronger observational
constraints on the cosmological parameters; in particular, combining
WMAP observations with other data sets gives extremely strong
evidence for non-zero cosmological constant. We have rerun our trees
for a LCDM cosmology with $\Lambda = 0.7$, $\Omega_{\rm m} = 0.3$, 
$h = 0.7$ and $\sigma_8 = 0.9$, a choice of parameters roughly consistent
with the analysis of the first year of WMAP data. After appropriate
scaling, we find substructure very similar to that in the SCDM
halos. Fig.\ \ref{fig:6}, for instance, shows that the average
cumulative mass function in LCDM haloes is very similar to the SCDM
mass function, although with 10--20 percent fewer objects at the
highest and lowest masses. The slightly reduced amplitude of the mass
function is consistent with numerical LCDM mass functions (e.g.\
Springel et al.\ 2001; Font et al.\ 2001; Governato et al.\ 2001;
Stoehr et al.\ 2002; Desai et al.\ 2004; Reed et al.\ 2004; De Lucia
et al.\ 2004; Gao et al.\ 2004b; Weller et al.\ 2004). A fit to the
slope of the LCDM mass function in the mass range 
$M_{\rm s}/M_{\rm vir, m} = 10^{-3}$--$10^{-4}$ gives 
$\alpha = (-0.9)$--$(-0.92)$.

Perhaps more importantly for our method, the pruning parameters (which
we determine by requiring self-consistency -- see paper I) derived for
the LCDM trees are very similar to those in the SCDM trees -- we find
$\overline{\Delta M} = 0.365$, $f_{\rm st} = 0.725$, and $n_{\rm o} = 2.3$ 
for model B in LCDM, versus $\overline{\Delta M} = 0.335$,
$f_{\rm st} = 0.677$, and $n_{\rm o} = 2.25$ for model B in SCDM. This
is encouraging, as it suggests the pruning parameters required for
self-consistency are fairly insensitive to the precise cosmology. We
will re-examine the dependence of halo substructure on cosmological
parameters in more detail in future work.

\section{Evolution of Subhalo Populations with Time}\label{sec:5}

In the previous section, we considered the properties of halo
substructure at the present-day, averaging over systems with very
different dynamical histories. The properties of individual subhaloes
will change systematically as they evolve within a larger system,
however, and thus the average properties of substructure in a halo
will depend on its dynamical age and assembly history. In this
section, we review how subhalo properties change with time, we show
how this dependence produces correlations between position, mass and
subhalo properties, and we quantify the relation between the assembly
history of a halo and the mean properties of its substructure.

\subsection{Properties of individual subhaloes as a function of their age}\label{subsec:5.1}

To characterise the age of each subhalo merging with the main system
we can define two different epochs within the merger tree, as
discussed in paper I; the `merger epoch' $z_{\rm m}$ when a subhalo
merges with the main trunk of the tree (that is the main progenitor of
the final system), and the `original merger epoch' $z_{\rm m,0}$ at
which it last exists as a distinct entity within the merger tree.
(These two redshifts are indicated schematically in Fig.\ 17 of paper
I.) The original merger epoch $z_{\rm m,0}$ will be greater than
$z_{\rm m}$ if the subhalo merges with another system before falling
into the main system, that is to say if it is a `higher-order'
subhalo in the sense defined in section \ref{sec:2}. The original
merger epoch is particularly important as it determines the original
density profile of the subhalo, before it is modified by tidal
stripping in subsequent mergers.

\begin{figure}
 \centerline{\psfig{figure=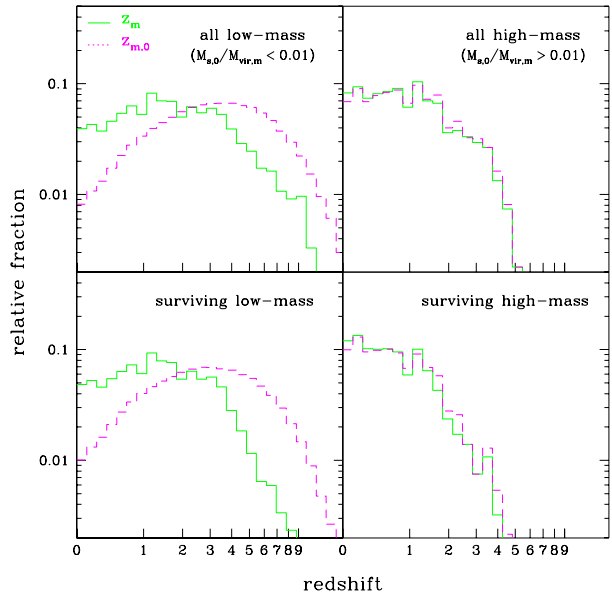,width=1.0\linewidth,clip=,angle=0}}
 \caption[]{Histograms of the merger epoch (solid line) and the original 
epoch at which systems first appeared in the merger tree (dashed line), 
for all subhaloes (top panel), and for those subhaloes that
survive to the present-day, in model B. Left and right-hand panels are
for low and high mass systems respectively.}
\label{fig:7}
\end{figure}

Fig.\ \ref{fig:7} shows the distribution of these two redshifts for
subhaloes in model B, divided up according to their mass when they
first merge with the main halo (left and right panels corresponding to
systems with original masses less than or greater than 1 percent of
the present-day mass of the main halo, respectively) and whether they
survive to the present-day. Most high-mass subhaloes have merged into
the main system since redshift 1--2, while low-mass subhaloes have
typical merger epochs of 1--4. For massive subhaloes, $z_{\rm m,0}$
is usually equal to $z_{\rm m}$, since most of these systems merge
directly with the main system. On the other hand the low-mass
subhaloes are often higher-order systems, and as a result the
distribution of $z_{\rm m,0}$ extends back to much higher redshift.

Recent numerical studies of halo substructure at very high resolution
have found similar results. Fig.\ 10 of De Lucia et al.\ (2004), for
instance, shows results comparable to the lower right-hand panel
(albeit for more massive haloes in a $\Lambda$CDM cosmology), while
Fig.\ 12 of Gao et al.\ (2004b) shows similar results for cluster,
group and galaxy haloes. In general, we predict more old substructure
than is seen in these simulations. In model B, about 30 percent of
surviving massive subhaloes merged before $z = 1$, whereas Gao et
al.\ find only 10 percent of their surviving systems are this old. We
will discuss this discrepancy further in paper III.

\begin{figure}
 \centerline{\psfig{figure=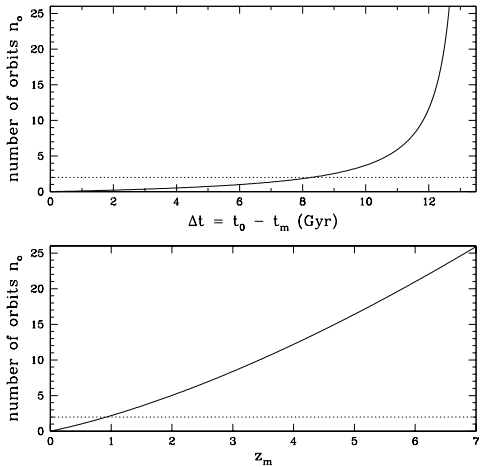,width=0.8\linewidth,clip=,angle=0}}
 \caption[]{The number of orbits (i.e.~pericentric passages) a
typical satellite has completed, as a function of the time it has
spent in the main system $\Delta_{\rm m} = t_{0} - t_{\rm m}$ (top
panel) or the merger epoch $z_{\rm m}$ (bottom panel).}
\label{fig:8}
\end{figure}

In paper I we also introduced the quantity 
$n_{\rm o} \equiv \Delta t/P_{\rm rad}$, the time a subhalo has spent 
in the main system in
units of the radial period at the virial radius of the main system at
the time the subhalo first fell in\footnote{$P_{\rm rad} \equiv 2\pi/\kappa$, 
where $\kappa = (v_{\rm c}/r_{\rm vir})(1 + {\rm d}\ln M/{\rm d}\ln r)^{1/2}$ 
is the epicyclic frequency at the virial
radius; see paper I.}, as an estimate of how many pericentric
passages it has undergone (pericentric passages occurring at 
$n_{\rm o} \sim$ 0.25--0.5, and once per radial period thereafter). 
Since the
radial period $P_{\rm rad}$ depends in a simple way on $z_{\rm m}$,
there is a straightforward relation between $n_{\rm o}$ and 
$z_{\rm m}$, as illustrated in Fig.\ \ref{fig:8}. In particular, for 
the SCDM cosmology considered here, $P_{\rm rad} \simeq 0.835 t_{\rm m}$,
where $t_{\rm m}$ is the age of the universe at redshift $z_{\rm m}$,
and the numerical factor depends slightly on the density profile of
the main system. Thus 
$n_{\rm o} = \Delta_{\rm m}/P_{\rm rad} \simeq (t_{0} - t_{\rm m})/0.835 t_{\rm m} \simeq 1.2 (t_{0}/t_{\rm m} - 1)$,
where $t_{0}$ is the present age of the universe.

Subhalo properties will depend on merger epoch for several reasons:
\begin{itemize}
 \item As the main system grows its satellite population extends to
progressively larger masses.
 \item As the main halo grows its virial radius increases, and
satellite orbits become correspondingly larger.
 \item The central densities and mean densities of infalling
satellites decrease with time, according to the EPS merger model and
the concentration-mass-redshift relation for haloes.
 \item The density profile of the main system changes with time,
although in practice this change is mainly confined to the outer
parts of the halo.
 \item Subhaloes will be stripped and disrupted with time, so the
properties of the surviving population will depend on how long they
have spent in the main system.
\end{itemize}
 
\begin{figure*}
 \centerline{\psfig{figure=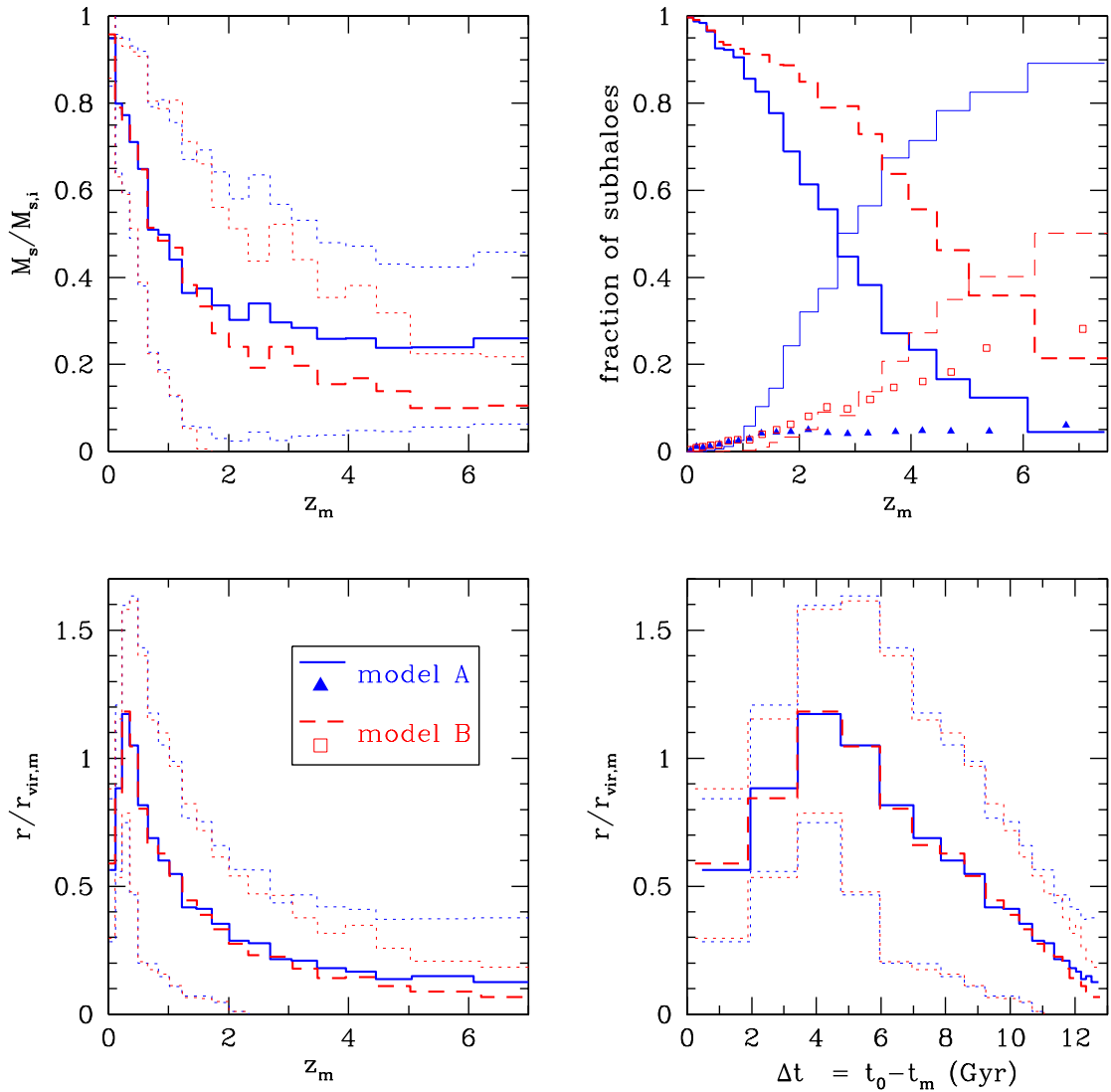,width=0.9\hsize,clip=,angle=0}}
 \caption[]{Subhalo properties as a function of merger epoch. Thick
solid and dashed lines indicate the mean values for models A and B
respectively; the dotted lines indicate the 1-$\sigma$ scatter in each
bin. The top-left hand panel shows the fraction of mass remaining.
The top right-hand panel shows the fraction of subhaloes that survive
(thick lines), that have been disrupted (thin lines), or that have
fallen in (points). The lower left-hand panel shows the radial
position relative to the virial radius versus merger epoch, and the
bottom right-hand panel shows the same quantity as a function of the
time elapsed since $z_{\rm m}$. }
\label{fig:9}
\end{figure*}

Fig.\ \ref{fig:9} shows the net effect of all these terms on subhalo
properties. The upper left-hand panel shows the average fraction of
their infall mass that subhaloes have retained as a function of merger
epoch. The thick lines show the mean value in each bin in merger
epoch, for model A (solid line) and model B (dashed line), while the
dotted lines show the 1-$\sigma$ scatter in each bin. The overall
behaviour is similar for both models; the mean value of $M/M_0$
decreases rapidly back to $z_{\rm m} \sim 2$, and is roughly constant
at higher redshifts, although the scatter around the mean remains
large, due to different subhalo orbits, masses and densities.

The top right-hand panel shows how the fraction of surviving subhaloes
decreases as a function of $z_{\rm m}$ (thick solid and dashed lines
for models A and B respectively), while the fraction of disrupted
subhaloes (thin solid or dashed lines) or of subhaloes that have
fallen into the centre of the main halo (solid triangles or open
squares, for models A and B respectively) increases progressively.

We can compare these results with the numerical results of Gao et al.\
(2004b, Fig.\ 14). In simulated clusters, their group finder can
locate 40--55 percent of all haloes accreted at $z_{\rm m} = 1$, and
only 10--20 percent of haloes accreted at $z_{\rm m} = 2$, whereas we
predict survival rates of 90 percent or 60--80 percent for these
redshifts. Some of the discrepancy is simply due to resolution; Gao et
al.\ require more than 10 bound particles to track a halo, which
eliminates those systems that retain less than 3--10 percent of their
infall mass (for the two mass ranges shown in their plot). Overall,
however, the numerical results find a systematically higher
disruption rate than we predict. They find that systems accreted at
$z_{\rm m} =$ 0.5 and 1 retain 40 percent and 25 of their infall mass
respectively, (Gao et al.\ 2004b, Fig.\ 13), whereas we predict the
same systems should retain 55 and 45 percent of their infall mass. We
will discuss the possible reasons for this disagreement further in
paper III.

The lower left-hand panel shows how subhaloes remain roughly
stratified within the main system, as a function of their merger epoch
(line styles are as in the top left-hand panel). Recently merged
systems are close to the ($z = 0$) virial radius, or may even be past
it as they go through the first apocentre of their orbit. Older
systems are found at progressively smaller radii, reflecting the size
of the main system when they first fell into it. In particular,
systems which merged before $z = 2$ and have survived to the present
day are typically located at 0.1--0.2 $r_{\rm vir,m}$. For the
average halo, $r_{\rm vir}(z)/r_{\rm vir}(z=0) = 0.4$ at $z = 1$, 0.22
at $z = 2$, 0.085 at $z = 4$, and 0.045 at $z = 6$, so in fact
surviving systems stay close to the virial radius of the halo at the
time they fell in. Finally, the lower right-hand panel shows the same
stratification in terms of the time elapsed since $z_{\rm m}$,
emphasising the synchronised pericentric passage of subhaloes accreted
within the past few Gyr, and the apocentric passage of systems
accreted 4--5 Gyr ago.

In summary, recently accreted subhaloes (e.g.\ those with 
$z_{\rm m} < 1$) are typically found at 0.5--1 $r_{\rm vir}$. 
Roughly 90 percent
have survived mass loss and tidal heating, and on average they retain
50--100 percent of the mass they had at infall. The oldest systems
(e.g.\ those with $z_{\rm m} > 2$, which fell in 11 Gyr ago) are
typically found at 0.1--0.2 $r_{\rm vir}$. Depending on which model
we consider, 40--90 percent of them survive, but they typically retain
only 20--30 percent of the mass they had at infall.

\subsection{Other correlations}\label{subsec:5.2}

\subsubsection{Correlations with mass}\label{subsubsec:5.2.1}

\begin{figure*}
 \centerline{\psfig{figure=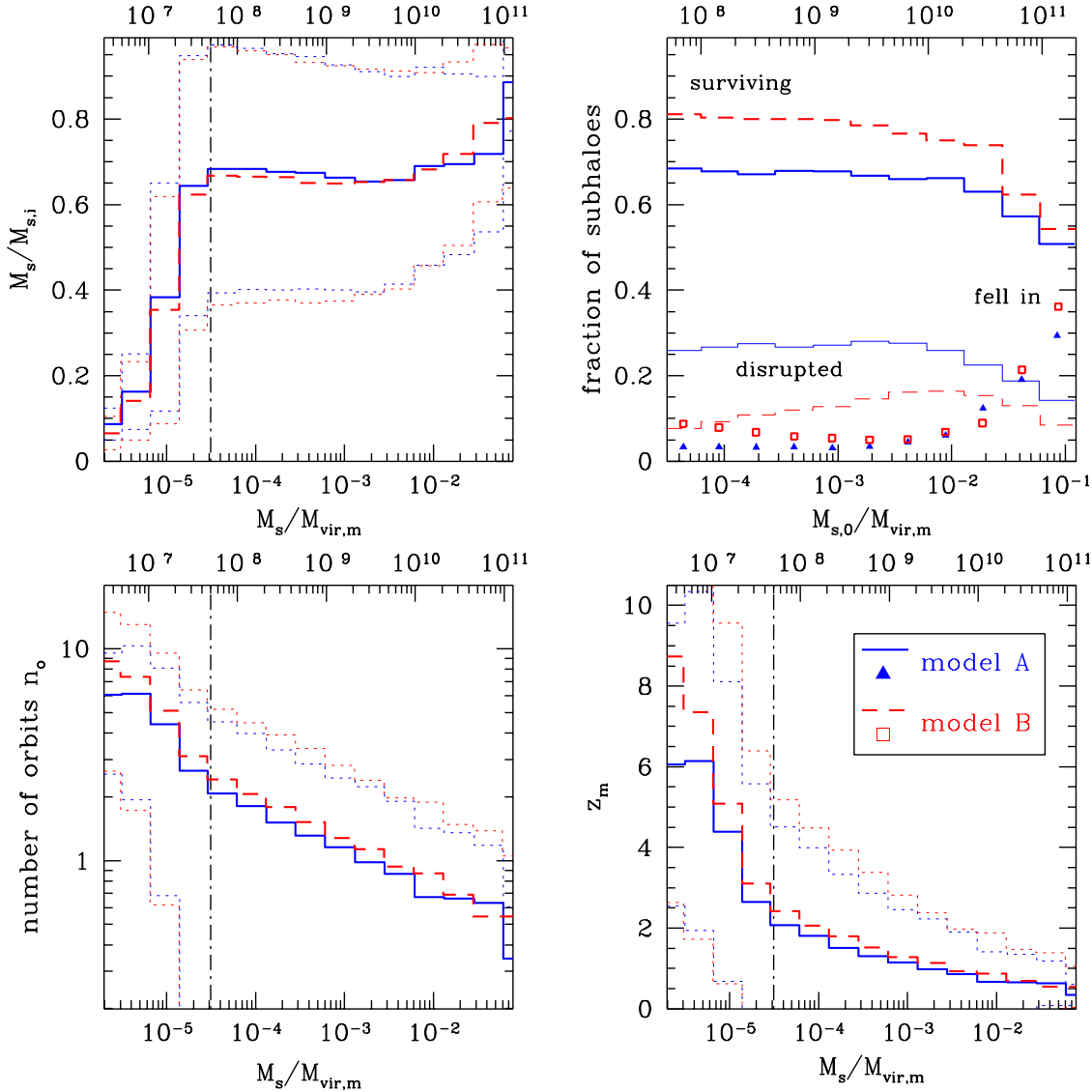,width=0.9\hsize,clip=,angle=0}}
 \caption[]{Subhalo properties as a function of mass. Thick solid and
dashed lines indicate the mean values for models A and B
respectively; the dotted lines indicate the 1-$\sigma$ scatter in each
bin. The top-left hand panel shows the fraction of mass remaining.
The top right-hand panel shows the fraction of subhaloes that survive
(thick lines), that have been disrupted (thin lines), or that have
fallen in (points). The bottom left-hand panel shows the average
number of orbits spent in the main halo, while the bottom right-hand
panel shows the average merger epoch. The dot-dashed line indicates
the mass resolution limit of the original merger trees.}
\label{fig:10}
\end{figure*}

The strong dependence of subhalo properties on merger epoch seen in
Fig.\ \ref{fig:9} introduces correlations between other subhalo
properties. Fig.\ \ref{fig:10}, for instance, shows how the degree of
stripping, the fractions of surviving, disrupted or centrally merged
subhaloes, the merger epoch, or the number of orbits spent in the
main halo (four panels, clockwise from top left) depend on the mass
of the subhalo. Line and point styles are as in Fig.\
\ref{fig:9}. The vertical dot-dashed line indicates the mass
resolution of the merger tree. Above this resolution limit, the
degree of stripping depends only weakly on mass (top left
panel). More specifically, the most massive subhaloes are
systematically younger and therefore less stripped, but otherwise the
average degree of stripping is almost independent of mass.

The fraction of surviving or disrupted systems is also roughly
independent of mass (top right panel), although dynamical frictions
causes a greater fraction of the massive subhaloes to fall into the
centre of the main system (points). On the other hand, the average
merger epoch (lower right) or equivalently the average number of
orbits spent in the main halo (lower left) are strongly correlated
with suhalo mass, tracking the increasing mass scale of the mergers
that form the main halo. Subhaloes below the resolution limit of the
merger trees (to the left of the dot-dashed lines) have mainly been
stripped, as indicated in the top left panel\footnote{To avoid
boundary effects at the mass resolution limit, the merger trees do
include some haloes below $5\times 10^{7}M_{\odot}$, but are very
incomplete below this limit.}. They are older than the average
subhaloes in that mass range would be if the merger tree extended to a
lower mass limit, as indicated by the break in the mean mass-redshift
relation in the bottom right panel.

Overall, massive systems have typically merged since a redshift of 1.
Roughly 10 percent have been disrupted by repeated mass loss, while
30--40 percent have been disrupted by falling into the centre of the
main system, such that only $\sim 50$ percent survive. Those that do
survive have only spent 1--2 orbits in the main system, and retain
70--90 percent of their infall mass. Low-mass systems are older
($z_{\rm m} \simgt 2$) and have spent up to 4--5 orbits in the main
system. 70--80 percent of them survive, and those that do retain 65
percent of their mass on average (with a 1-$\sigma$ scatter of
0.35--0.95 for individual systems). As before, numerical simulations
generally find slightly higher rates of mass loss and disruption. De
Lucia et al.\ (2004, Fig.\ 11), for instance, find that their
subhaloes retain only 40--50 percent of their infall mass on average.
Nagai \& Kravtsov (2004) also find a somewhat lower value for the
retained mass fraction as a function of radius (cf.\ their Fig.\ 4
versus the top left-hand panel of Fig.\ \ref{fig:10}), but their
values are expressed in terms of the maximum mass a subhalo had at any
point in its merger history (essentially our `original mass'), and are
thus hard to compare directly. Given a mean mass loss rate of 
$\sim 20$--30 percent for systems before they enter the main system, our
results appear to be similar to theirs. On the other hand, Gill et
al.\ (2004b) find similar results for the number of orbits fairly
massive satellites have spent in the main system (see their Fig.\
6). We will compare our results with simulations in more detail in
paper III.

\subsubsection{Correlations with position}\label{subsubsec:5.2.2}

\begin{figure*}
 \centerline{\psfig{figure=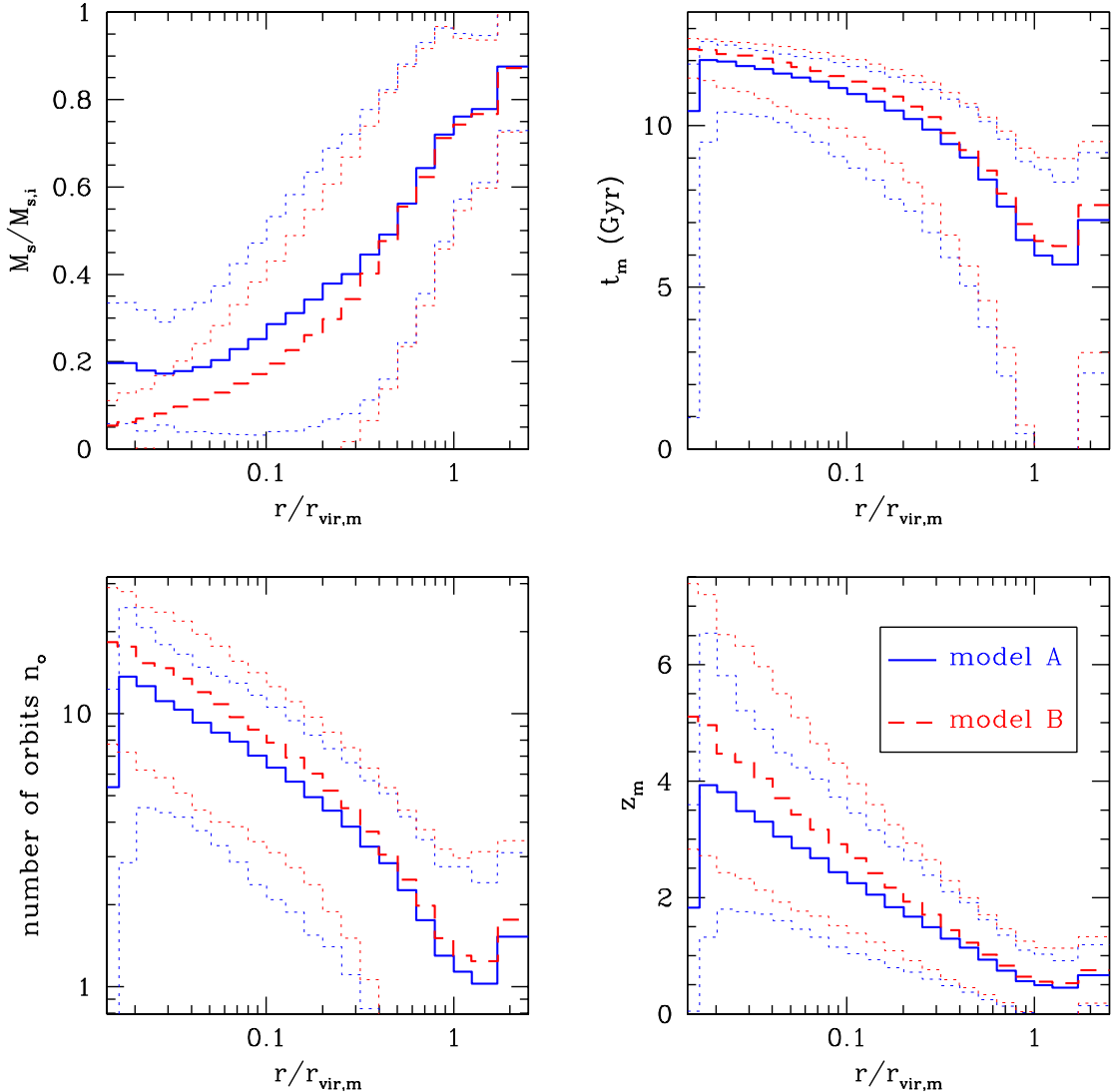,width=0.9\hsize,clip=,angle=0}}
 \caption[]{Subhalo Properties as a function of position. Line styles
are as in Figs.\ \ref{fig:9} and \ref{fig:10}. The four panels show
the average and scatter in the degree of stripping (top left), the
merger time (top right), the number of orbits spent in the main system
(bottom left) and the merger epoch (bottom right), in different radial
bins.}
\label{fig:11}
\end{figure*}

Finally, the increasing size of satellite orbits as a function of time
will produce correlations between radial position in the halo and
other subhalo properties. Fig.\ \ref{fig:11} shows the average degree
of stripping, the average merger time, the average merger epoch, and
the number of orbits subhaloes have spent in the main system as a
function of radial position (four panels, clockwise from upper left).
Line styles are as in Figs.\ \ref{fig:9} and \ref{fig:10}. These
distributions have interesting implications for the dynamical
evolution of the central regions of haloes. They indicate, for
instance, that subhaloes in the central 2--3 percent of the virial
radius (roughly 6--10\,kpc, for the Milky Way) have been there for
10--12 Gyr (or since a redshift of 2--6), completing tens of orbits
and losing 90 percent of their infall mass through tidal stripping. We
also note that the difference between models A and B increases for
central subhaloes; in model A the average surviving mass fraction is
constant and equal to $\sim$ 20 percent in the central 5 percent of
the halo, while in model B it continues to decrease at small radii,
reaching 5 percent at 0.02 $r_{\rm vir,m}$. (The sudden change in the
distributions at 0.02 $r_{\rm vir,m}$ occurs because we have
considered systems that pass within 0.01 $r_{\rm vir,m}$ of the centre
of the main halo to have fallen in and been disrupted, as explained in
section \ref{sec:2}.)

On the other hand, subhaloes close to $r_{\rm vir,m}$ have typically
merged with the main system at $z_{\rm m} < 1$ and have completed only
2--3 orbits or less. They have generally been in the main system less
than 6--7 Gyr, and retain 60--90 percent of their mass. At these large
radii, the results of models A and B are very similar. Gao et al.\
(2004b) find similar results for subhaloes close to the virial radius
in simulations (cf.\ their Fig.\ 15), but at smaller radii there
systems are slightly more stripped and have slightly lower merger
epochs.

\section{Global age indicators}\label{sec:6}

Having determined how the properties of individual subhaloes relate to
their age and position with a halo, we will now examine how the
cumulative distributions of subhalo properties depend on the merger
history of their parent halo. As highlighted recently by Diemand et
al.\ (2004c), cumulative mass functions of logarithmic slope $-1$ are
invariant in simple merger scenarios where all (or at least a fixed
fraction, independent of mass) of the subhaloes in the progenitors
survive in the final system. If 10 systems of mass $M_0$ each with
one subhalo of mass $10^{-2}\,M_0$ merge together, for instance, the
result is a system of mass $M_1$ with ten subhaloes of mass
$10^{-3}\,M_1$, and the slope of the mass function will be unchanged.

Even if we assume that the mass function does have this form in  some
initial population of haloes, subsequent mergers between these haloes
may change its shape for several reasons. First, dynamical friction
will cause subhaloes with more than a few percent of the mass of
their parent to fall into the centre of its potential and be
disrupted after a few orbits, truncating the mass function at an
upper limit which depends on the age of the system. Accretion or
mergers with very small haloes will increase the mass of the main
system without adding to its massive substructure, shifting or
steepening the mass function. Finally, if a halo is isolated and
grows relatively little over some period, tidal mass loss will
systematically reduce the mass of its subhaloes, decreasing the
amplitude of the mass function.

In summary, the subhalo population evolves through two competing
effects: injection and mass loss or disruption. The former occurs
only during mergers, while the latter operates at a rate that depends
on subhalo mass and on the orbital period of the main system, or
equivalently on cosmic time. Thus we expect to see systematic trends
in the relative mass function of a halo that correlate with its
assembly history. In particular, we expect the amplitude of the
relative mass function to be largest when a halo is dynamically
`young' (in the sense that it has assembled a large fraction of its
mass recently), and smallest when it is dynamically `old', since the
subhalo population will gradually be stripped and disrupted during
the quiescent periods in a halos evolution. In this sense `young' and
`old' halos are analogous to relaxed and unrelaxed galaxy clusters;
indeed observations of group and cluster luminosity functions may
provide a simple observational test of the evolutionary history and
formation rate for haloes of different mass (e.g.\ Jones et al.\
2003; see also Trentham \& Tully 2002; Roberts et al.\ 2004; D'Onghia
\& Lake 2004).

\begin{figure}
 \centerline{\psfig{figure=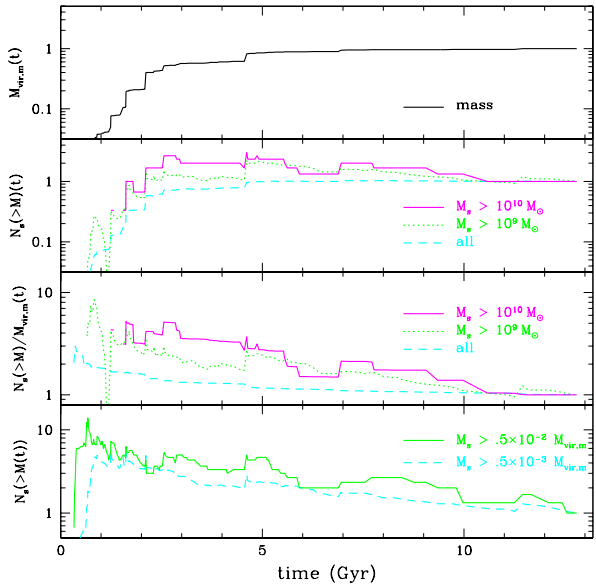,width=1.0\linewidth,clip=,angle=0}}
 \caption[]{(Top panel) The total mass of a particular halo, relative
to the mass at $z = 0$. (Second panel) The number of subhaloes in the
same system over some fixed mass limit, relative to the number at the
present day. The dashed, dotted and solid lines are for all
subhaloes, those more massive than $10^{9} M_{\odot}$, and those more
massive than $10^{10} M_{\odot}$ respectively. (Third panel) the
number of subhaloes per unit mass of the main halo, relative to the
same quantity at the present day. The mass bins are as in the second
panel. (Bottom panel) The number of subhaloes with more than some
fixed fraction of the mass of the main halo at that time, relative to
the same quantities at the present day.}
\label{fig:12}
\end{figure}

Fig.\ \ref{fig:12} shows how the number of subhaloes in one particular
halo is related to its mass accretion history. The top panel shows the
mass of the system as a function of time, $M_{\rm m}(t)$, normalised
to 1 at the present day. The second panel shows the number of
subhaloes in the system over a fixed mass threshold of 
$10^{10} M_{\odot}$ (solid line) or $10^{9} M_{\odot}$ (dotted line), as 
well as the total number down to the resolution limit. In each case we have
normalised to the number over the same threshold at the present
day. Although the patterns are slightly obscured by the small number
of massive subhaloes, we see that while the total number of subhaloes
tracks the relative mass of the system, the number of massive
subhaloes peaks after every major merger. There are two to three
times more subhaloes over $10^{10} M_{\odot}$ after every major merger
in the last 10 Gyr, for instance (solid line). After each merger the
number then decreases over 1--2 Gyr, as these subhaloes fall in and
are disrupted, or are tidally stripped to the point that they drop
below the mass threshold. Comparing the last and second-to-last
peaks, there is also a marginal indication that the timescale for this
decline increases with time, as we would expect given its scaling with
orbital timescales, which in turn scale roughly as the age of the
universe (see section \ref{subsec:5.1}). The dotted line shows
similar patterns for slightly less massive subhaloes, although perhaps
with a reduced amplitude at early times, and a more gradual decay.

Of course the mass of the main system is changing throughout these
events, complicating the interpretation of their effect on the
relative mass function. The third panel from the top shows the same
quantities as in the second panel, divided by the mass of the main
system at that time, and normalised to one at the present day. (In
other words, the quantity plotted is the number of subhaloes per unit
mass of the main halo at that time that exceed a fixed mass
threshold, in units where the present-day value is 1.) The trends
described above are now clearer, and we see that even for the lowest
mass threshold (dashed line), the number of subhaloes per unit mass
(or equivalently the amplitude of the relative mass function) is 2--3
times larger just after the halo has assembled.

The change in the relative mass function at a given relative, rather
than fixed, mass is shown in the bottom panel. For the most massive
systems ($0.5\times 10^{-2} M_{\rm vir,m}$, corresponding to the
$\sim$5 most massive systems at the present day), the amplitude of
the relative mass function is an order of magnitude larger at early
times, and even for a lower mass threshold ($0.5\times 10^{-3} M_{\rm vir,m}$,
corresponding to the $\sim$50 most massive systems at the
present day), the amplitude is 4--5 times the present-day value at
early times. Overall, we conclude that systems which have experienced
recent major mergers should have more massive substructure, but that
many of these new subhaloes will be destroyed after a few dynamical
times, particularly if they are very massive, consistent with the
results of paper I. The dependence of substructure on the
evolutionary history of the parent halo has been studied recently in
numerical simulations by Gill et al.\ (2004b), who find a similar
decay in the number of massive satellites after systems have
assembled (e.g.\ their Fig.\ 3).

\begin{figure}
 \centerline{\psfig{figure=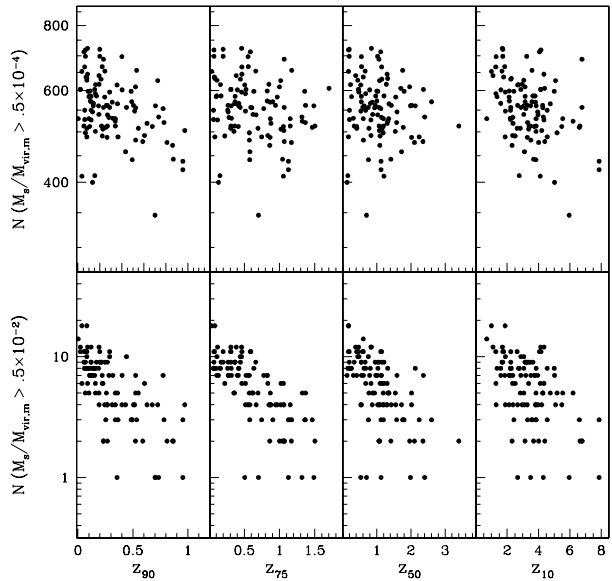,width=1.0\linewidth,clip=,angle=0}}
 \caption[]{The number of subhaloes with more than some fraction of
the mass of the main system (at the present day), as a function of
the formation epoch of the halo. }
\label{fig:13}
\end{figure}

These patterns also suggest that we look for correlations between the
amount of substructure in a halo at $z = 0$ and its formation epoch.
Fig.\ \ref{fig:13} shows the number of subhaloes more massive that
$0.5\times 10^{-2}$ or $0.5\times 10^{-4}$ times the mass of the main
system $M_{\rm m}$, as a function of the formation epochs defined in
paper I, that is the epochs by which the main progenitor of the system
had built up 90, 75, 50 or 10 percent of its final mass\footnote{Note
that while the expected distribution of formation epochs $z_{f}$
cannot be calculated analytically for $f < 0.5$, as a tree may have
more than one progenitor of this mass, it remains well defined if we
always choose to follow the most massive branch at each branching of
the tree, and define $z_{f}$ as the time when this branch reaches a
fraction $f$ of the total halo mass at $z = 0$.}. The number of
massive systems is strongly correlated with the recent merger
history, as indicated by $z_{90}$, $z_{75}$ or $z_{50}$. As expected
from the preceding discussion, younger systems have more massive
subhaloes, the mean number varying by a roughly factor of five over
the range of redshifts sampled. The total number of haloes down to
$0.5\times 10^{-4} M_{\rm vir, m}$ is also correlated with the recent
merger history of the system, or with its overall age, as indicated
by $z_{10}$ for instance, although the correlation is weaker and the
overall variation in number in is less than a factor of two.

Similar correlations for $z_{50}$ and $z_{25}$ were reported recently
by Gao et al.\ (2004b, Fig.\ 8), who also pointed out that the scatter
in the correlation may be somewhat larger for $z_{75}$ or $z_{90}$, as
these redshifts can be low even for haloes formed by recent major
mergers of relatively old systems with little internal
substructure. We do find somewhat more scatter in the results for
$z_{90}$ than for $z_{75}$ or $z_{50}$, although the trend is not very
strong.

\begin{figure}
 \centerline{\psfig{figure=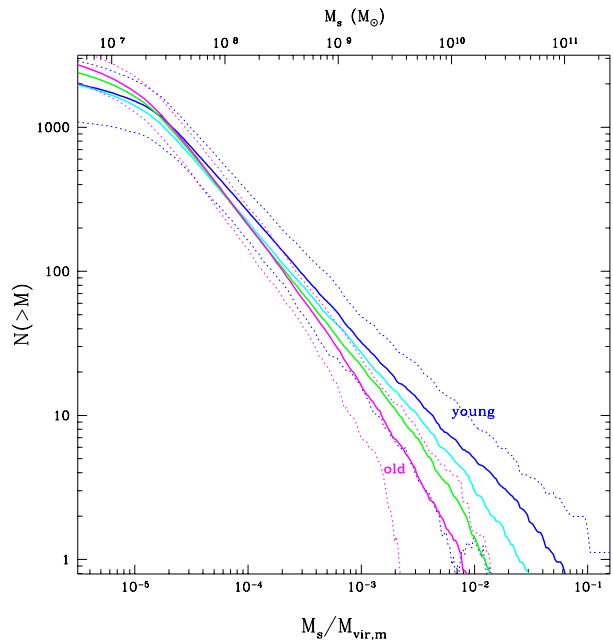,width=1.0\linewidth,clip=,angle=0}}
 \caption[]{Cumulative mass functions for systems binned by formation
epoch $z_{75}$. The four solid lines are for $z_{75} < 0.3$, 
$z_{75} =0.3$--0.55, $z_{75} = 0.55$--1.00, and $z_{75} > 1.00$ from right 
to left at the high mass end. The thin dotted lines show the 2-$\sigma$
halo-to-halo scatter for haloes in the first and last bins.}
\label{fig:14}
\end{figure}

These correlations produce systematic changes in the cumulative mass
and velocity distributions as a function of formation epoch. Fig.\
\ref{fig:14} shows the cumulative mass functions for systems binned by
formation epoch $z_{75}$. The four bins: $z_{75} < 0.3$, $z_{75} =0.3$--0.55, 
$z_{75} = 0.55$--1.00, and $z_{75} > 1.00$, were chosen to
produce roughly equal numbers of haloes in each bin. The effect seen
in Fig.\ \ref{fig:12} is visible here as well; old systems have a
steeper mass function that is more strongly truncated at the high
mass end, relative to the younger systems. (We note that the overall
scatter in the mass of the most massive satellite shown in Fig.\
\ref{fig:14} is very similar to that found by De Lucia et al.\ 2004
for haloes of a similar mass -- cf. their Fig.\ 4.) In the mass range
$M_{\rm s}/M_{\rm vir, m} = 10^{-3}$--$10^{-4}$, the mean slope of the
cumulative mass function is -0.89, -0.92, -0.98 and -1.09 for the
four bins respectively. As expected from Fig.\ \ref{fig:13}, there is
only a slight change in the amplitude of the mass function at the low
mass end, although this might be stronger if we had binned the haloes
by $z_{50}$ or $z_{10}$. Overall, the systematic dependence of the
mass function on dynamical age may explain some of the variation in
cumulative mass and velocity functions reported in recent simulations
(e.g.\ Desai et al.\ 2004; De Lucia et al.\ 2004; Gao et al.\ 2004b).

\begin{figure}
 \centerline{\psfig{figure=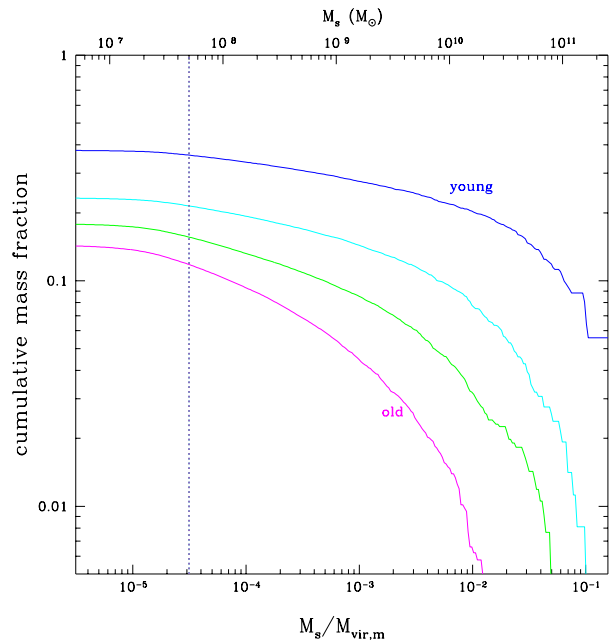,width=1.0\linewidth,clip=,angle=0}}
 \caption[]{The fraction of the mass within the virial radius
contained in subhaloes of mass $M$ or larger. The subhaloes have been
binned by formation epoch $z_{75}$, as in Fig.\ \ref{fig:14}, with age
increasing from top to bottom. }
\label{fig:15}
\end{figure}

The systematic change in the mass function is particularly important
when calculating the fraction of the halo mass contained in
substructure, since it is dominated by massive subhaloes. Fig.\
\ref{fig:15} shows how this quantity varies by a factor of 4 over the
age bins considered above. As we discuss in paper III,
the systematic dependence of the amount of substructure on the
dynamical history of the halo may account for part of the discrepancy
between numerical substructure mass functions and the semi-analytic
predictions. Only very recently have high-resolution results been
available for sufficiently many systems to test for halo-to-halo
variations. Fig.\ 7 of Gao et al.\ (2004b) or Fig.\ 1 of Gill et al.\
(2004b) show how much the mass fraction in substructure varies from
halo to halo, for instance, in high-resolution $\Lambda$CDM
simulations. The scatter is similar to that shown in Fig.\
\ref{fig:15}, although Gao et al.\ (2004b) finds a substantially
lower mean fraction, whereas Gill et al.\ (2004b), using an adaptive
group finder, finds results closer to those shown here. Finally we
note that the scatter in the mass fraction from halo-to-halo can be
very large, as a few of our systems have subcomponents with a
substantial fraction (20--30 percent) of the total halo mass (this
also explains why the average cumulative mass fraction in Fig.\
\ref{fig:15} is already substantial at large masses.

\section{Grouping and Encounters}\label{sec:7}

\subsection{Modelling group dynamics}\label{subsec:7.1}

As discussed in paper I, our semi-analytic model contains higher-order
information about substructure merging into the main system. As a
result of the pruning process, we can identify in the final merger
tree groups of subhaloes which should be kinematically associated 
with one-another when they first fall in, because they represent
substructure from a single merging system. These kinematic groups 
are assigned a mean orbit, and each member of the group is given 
an initial offset in position and in velocity relative to this mean, 
as described in paper I. In this way we reproduce approximately the
patterns found in spatially resolved simulations of halo mergers.

\begin{figure}
 \centerline{\psfig{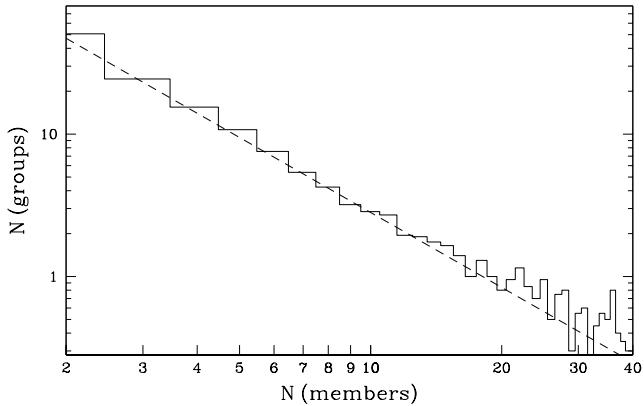}}
 \caption[]{The average number of kinematic groups per halo with a
given number of members. The dashed line has a slope of -1.75.}
\label{fig:16}
\end{figure}

An average system contains many kinematic groups, although most have
only a few members. Fig.\ \ref{fig:16} shows a histogram of the
average number of groups versus the number of members in the
group. It is roughly a power law of slope -1.75, with on the order of
a hundred groups in the average halo, including one or two large
groups with 10 members or more.

\begin{figure}
 \centerline{\psfig{figure=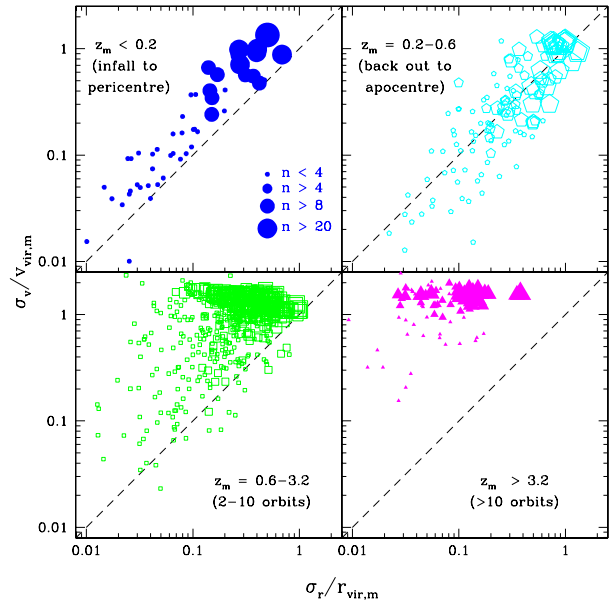,width=1.0\linewidth,clip=,angle=0}}
 \caption[]{The spatial dispersion and velocity dispersion of
kinematic groups in a set of haloes (at $z = 0$). The size of the
symbol shows the number of members, as indicated in the first
panel. Through phase mixing and subsequent virialisation,
successively older groups have moved to the right (top two panels)
and then to the top of the plot (bottom two panels).}
\label{fig:17}
\end{figure}

As groups fall in and evolve in the main system, their members will be
scattered and disrupted, particularly since we do not include the
self-gravity of a group in our orbital calculations (except in the
approximate sense that the haloes which are most tightly bound to
their parent system are subsumed in the parent during the pruning of
the merger tree). We can study the evolution of groups with time by
plotting the spatial dispersion and velocity dispersion of groups as
a function of their age Fig.\ \ref{fig:17} shows these dispersions,
normalised to the virial radius $r_{\rm vir,m}$ and the virial
velocity $v_{\rm vir,m}$ of the main system, at $z = 0$. For each
group, the size of the symbol indicates the number of members (which
is roughly correlated with the total group mass), while the four
panels show groups with successively earlier infall times,
corresponding to systems which are falling in for the first time (top
left), systems which have passed through pericentre once (top right),
systems which have completed 2--10 orbits (lower left) and systems
which have completed more than 10 orbits (lower right).

In our model for grouping, groups start out with 
$\sigma_{\rm v}/v_{\rm c} \simeq \sigma_{\rm r}/r_{\rm vir,m}$, 
each dispersion
being proportional to the cube root of the mass of the group (see
paper I, section 5.3). This corresponds to the dashed line running
diagonally across each panel. The four panels illustrate several
phases in the subsequent evolution of group dynamics. By the time of
their first pericentric passage, recently-merged groups have been
compressed and heated slightly such that 
$\sigma_{\rm r}/r_{\rm vir,m} < \sigma_{\rm v}/v_{\rm c}$, 
and thus they lie to the left of
the dashed line (top left panel). As they continue to orbit,
conservation of phase-space density turns the scatter in velocities
into a scatter in positions, so the groups move back towards the
fiducial line. Slightly older groups (top right panel) continue to
experience this orbital heating, moving them upwards in the plot. As
groups evolve through many orbits, (lower left panel) their velocity
dispersion increases until it reaches the velocity dispersion of the
main system; at this point, any information about orbital
correlations within the group has effectively been lost. The members
of old groups have systematically smaller orbits, however, as
discussed in paper I. Thus the spatial dispersion of the oldest
groups (lower right panel) is smaller relative to virial radius of
the main system at the present day.

\subsection{Encounters and harassment}\label{subsec:7.2}

While we cannot treat interactions between subhaloes realistically in
our model, we can get an indication of their relative importance by
flagging close encounters between subhaloes, that might lead to
orbital scattering, mass loss or disruption. The number of encounters
recorded will depend strongly on the maximum impact parameter 
$b_{\rm max}$ used to define a close encounter. We choose the impact
parameter such that the close encounters we record have roughly the
same tidal effect as a pericentric passage in the subhalo's
orbit. This corresponds to choosing $b_{\rm max}$ such that the mean
density of the larger member of the interacting pair interior to
$b_{\rm max}$ is comparable to the density of the main system
interior to the pericentre of a typical subhalo orbit. Assuming
self-similarity between the inner regions of the subhaloes and the
inner part of the main system, this will be the case for encounters at
impact parameters less than a few times the peak circular velocity
radius of the larger subhalo, $r_{\rm p}$. Thus it is useful to
define encounter statistics in terms of the scaled impact parameter
$x \equiv b/r_{\rm p}$, where $r_{\rm p}$ is the peak radius of the
more massive halo in an interacting pair.

\begin{figure}
 \centerline{\psfig{figure=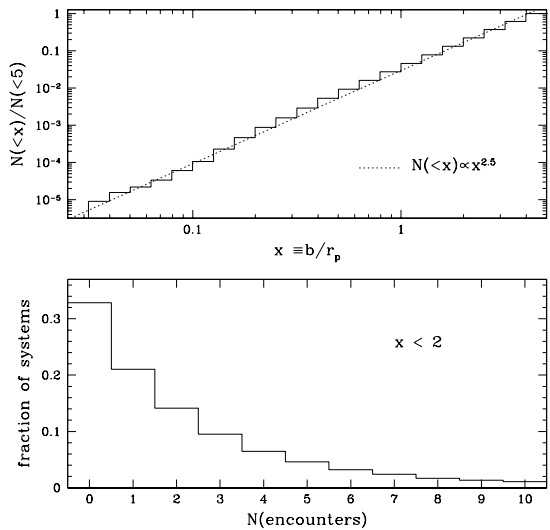,width=0.9\linewidth,clip=,angle=0}}
 \caption[]{(Top panel) the number of close encounters between haloes
with impact parameters $b$ less than $x$ times the peak radius of the
larger system, $r_p$. The cumulative number goes roughly as $x^{2.5}$,
as indicated by the dotted line. (Bottom panel) the fraction systems
that have had a given number of encounters with systems of equal or
greater mass, at impact parameters of less than twice the peak radius
of the more massive system.}
\label{fig:18}
\end{figure}

The top panel of Fig.\ \ref{fig:18} shows how the number of encounters
varies with $x$. We have normalised the number of encounters by the
total number for $x < 5$. The cumulative number is roughly a
power-law: $N(<x) = N_o\,x^{2.5}$. We expect encounters with more
massive systems at $x \simlt 2$ (that is encounters where the haloes
overlap at or within each other's peak radii) to heat and strip
subhaloes roughly the same way pericentric passages do. In what
follows we will refer to such events as `major' encounters. Overall,
find that more than 65 percent of subhaloes have had at least one
major encounter defined in this way, and 45 percent have had more
than one. The bottom panel of Fig.\ \ref{fig:18} shows a histogram of
the fraction of systems that have had a given number of major
encounters. 

We note that subhalo encounter rates have also been measured in 
simulations by Tormen et al.\ (1998), and more recently by Knebe et 
al.\ (2004). While both of these studies find that encounters are common
(affecting $\sim$ 60--75 or 30 percent of all systems in the two studies, 
respectively), it is hard to compare their measured rates directly with 
ours as each study uses quite different criteria to define an encounter.

\begin{figure}
 \centerline{\psfig{figure=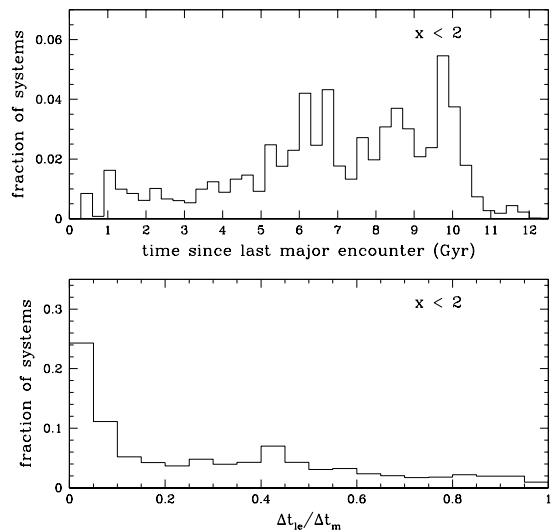,width=0.9\linewidth,clip=,angle=0}}
 \caption[]{The distribution of times since the last major encounter
occurred (top panel), and the time delay between infall and the last
encounter $\Delta t_{\rm le}$ as a fraction of the total length of
time the subhalo has been in the main system, $\Delta t_{\rm m}$
(bottom panel).}
\label{fig:19}
\end{figure}

Fig.\ \ref{fig:19} shows the distribution of times at which major
encounters occurred (top panel), and the delay between infall and the
last major encounter as a fraction of the total length of time the
subhalo has been in the main system (bottom panel). Most major
encounters occur at fairly early times, and they also usually occur
shortly after the subhalo in question first merged into the main
system. This is probably because the density of subhaloes within a
halo is highest at early times or after a recent period of
accelerated growth, as discussed in section \ref{sec:6}. In
particular, collisions and interactions between members of a kinematic
group are most likely to occur at the first pericentric passage of the
group, when the orbits of its members are focused together.
Triggering through collisions, encounters and pericentric passages may
play an important role in determining the star-formation histories of
the dwarf galaxies of the Local Group, as discussed in Taylor (2002).
We will consider this point in more detail in future work.

Since a large fraction of our subhaloes will have had a close
encounter with another system of comparable mass by $z = 0$, we may be
underestimating the extent to which they are tidally heated or
stripped. In clusters, the dynamical effect of multiple encounters
has been referred to as `harassment' (Moore et al.\ 1996). We can
quantify the possible effect of harassment on the subhalo mass
function by stripping from a subhalo the average fraction of its mass
it would lose in a pericentric passage, every time it experiences an
encounter with a more massive system at an impact parameter of
$b_{\rm max} = 2\,r_{\rm p}$ or less. The motivation for this
approximation, as mentioned above, is that the tidal force generated
by the two events should be comparable. In the bottom right-hand
panel of Fig.\ \ref{fig:5}, we showed the effect of this additional
term on the differential mass function. The curve labelled
`collisions' corresponds to a case where every subhalo experiencing a
collision with a larger subhalo at $x < 1$ is reduced to a fraction
$f_{\rm st} = 0.73$ of its mass, as if it had experienced another
episode of stripping in the merger tree (see section \ref{sec:2}). In
the case labelled `extreme collisions', satellites are reduced to 0.1
$f_{\rm st}$ after every major encounter, leading to immediate
disruption of the subhalo in most cases.  Overall, the net effect of
harassment is fairly small -- in the former more realistic case it is
equivalent to the average subhalo losing an extra 10 to 20 percent of
its mass -- but since this effect is systematic and probably would
occur in a real system, it should be kept in mind when we
compare our results with numerical simulations below.

\section{Discussion}\label{sec:8}

In this paper, we have discussed the basic properties of halo substructure,
as predicted by the semi-analytic model of halo formation described in
paper I. This model includes several distinct components:
\begin{itemize}
 \item Halo merger histories are generated randomly, using the
merger-tree algorithm of Somerville and Kolatt (1999).

 \item The higher order branchings in these trees are `pruned', using
the method described in paper I, to determine whether each branch
merging with the main trunk contributes a single subhalo 
or a group of associated subhaloes to the main system. This produces
a single list of subhaloes merging with the main system at various 
redshifts.

 \item Each subhalo from this final list is the placed on a random
orbit starting at the virial radius of the main system, and evolved
using the analytic model of satellite dynamics described in TB01,
experiencing orbital decay due to dynamical friction, and heating and
stripping due to tidal forces.

 \item Haloes which were associated with a given parent before
pruning fall in together with the parent on similar orbits, as part of
a kinematic group.

 \item The properties of the main system change over time, its mass
growing according to the merger tree and its concentration changing
according to the relations in ENS01.

 \item No baryonic component is included in the models presented here,
although one can easily be added, given a prescription for gas cooling
and star formation.
\end{itemize}

Overall accuracy and parametric dependence are important considerations
for any semi-analytic model. From paper I, we expect our dynamical model 
of satellite evolution to predict the properties of individual subhaloes 
to an accuracy of roughly 20 percent until they have lost a large fraction 
($\simgt$ 80--90 percent) of their original mass. This component of the 
full semi-analytic model has three principal free parameters: the (halo) 
Coulomb logarithm, $\Lambda_{\rm h}$, which determines the strength of 
dynamical friction, the heating coefficient $\epsilon_{\rm h}$, which 
determines the mass loss rate, and the disruption criterion $f_{\rm crit}$, 
which determines how long stripped systems continue to survive as bound 
objects. The values of these parameters were fixed by comparison with the 
simulations of Velazquez and White (1999) and H03, as described in TB01 and 
in paper I. 

Examining the dependence of our results on the values of these 
parameters, (in section \ref{sec:4}), we conclude that $\Lambda_{\rm h}$ 
is probably well-enough determined for most applications. The heating 
parameter $\epsilon_{\rm h}$ may introduce larger uncertainties, as it 
was determined from a limited number of simulations with $10^3$--$10^4$
particles each. As a result it may be that we {\it over}estimate the 
importance of heating somewhat, and subhalo masses are $\sim$ 20--30 percent 
larger than we predict. The disruption criterion is also slightly uncertain, 
although it only affects the survival of highly stripped subhaloes. Here 
again, we are likely to overestimate the disruption rate, as we are basing 
our results on simulations of finite resolution.

The rest of the semi-analytic model has no other free parameters, 
although there are a number of other minor uncertainties, assumptions 
and approximations in the model. The main assumptions concern the form 
of the density profile for the main halo, the dependence of halo 
concentration on mass and redshift, and the distribution of orbital 
parameters when haloes first merge. The average halo density profile 
is probably well-enough determined from simulations for most of the 
applications considered here, as is the initial distribution of satellite 
orbits. The halo concentrations used here (from ENS01) may be somewhat 
too small for low-mass haloes (cf.\ Bullock et al.\ 2001); in this case 
we are probably underestimating subhalo masses by 10--15 percent.

The model also makes a number of simplifications or approximations, 
treating the main halo as spherically symmetric, for instance, and 
ignoring some of the higher-order terms in subhalo evolution, such 
as collisions between systems or `harassment'. The anisotropy of 
subhalo orbits has been studied by several authors in recent 
simulations (e.g.\ Knebe et al.\ 2004; Gill et al.\ 2004b;
Aubert, Pichon \& Colombi 2004; Benson 2004) and is generally
fairly small. We do not expect it to affect our results substantially. 
Similarly, while the density profiles of simulated CDM haloes are 
flattened or triaxial to a certain degree, the corresponding potentials 
are only slightly non-spherical, so we do not expect that flattening 
will change our results on the spatial distribution of satellites 
substantially. The higher-order terms in subhalo evolution probably
increase mass loss and disruption rates slightly, but the estimates of 
section \ref{sec:4} suggest this is at most a 10--20 percent effect.

In summary, given the calibration performed in TB01 and paper I, 
our semi-analytic model has no free parameters, although it includes
various assumptions that could be modified, and various simplifications 
that could be treated in more detail. We expect it to be accurate to
roughly 20 percent, the remaining inaccuracy begin due to several
competing effects, including possible overestimates of heating and 
disruption rates, or underestimates of subhalo concentration and mass loss 
due to encounters. Given this estimated accuracy, it is interesting to 
compare our results with self-consistent simulations of halo formation.

Overall, the semi-analytic model predicts simple, universal distributions 
of subhalo mass, circular velocity, and position within the main halo, 
similar to those seen in simulations. We also predict strong correlations 
between a subhalo's merger epoch or formation epoch and its properties, 
and as a result, between the dynamical history of a halo and the average 
properties of its substructure. Here again, these are qualitatively similar 
to trends seen in recent studies of substructure in sets of simulated haloes.

In paper III, we will make a more quantitative comparison between numerical 
and semi-analytic predictions. We have seen preliminary indications of the 
results of this comparison in section \ref{sec:5}: subhaloes in simulations 
generally experience more mass loss and are disrupted faster than in the 
semi-analytic model. While this disagreement could be due to a systematic 
error in the semi-analytic model, it could also indicate that the simulations 
are still subject to important resolution effects. 
We will discuss this issue in more detail in paper III.

\section*{Acknowledgements}

The authors wish to thank E. Hayashi, S. Ghigna, B. Moore, J. Navarro
and T. Quinn for providing data from their simulations for comparison
with our model. We also wish to thank E. Hayashi, T.\ Kolatt, A.\
Kravtsov, J.\ Navarro, J. Silk, and S.\ White for helpful discussions.
JET gratefully acknowledges the support of a postgraduate scholarship
from the Natural Sciences \& Engineering Research Council of Canada
(NSERC) during the initial stages of this work, and support from the
Leverhulme Trust and from the UK Particle Physics and Astronomy
Research Council (PPARC) in the latter stages. AB gratefully
acknowledges support from NSERC through the Discovery and the
Collaborative Research Opportunities (CRO) grant programs.


\end{document}